\documentclass{article}

\usepackage{arxiv}

\usepackage[utf8]{inputenc} 
\usepackage[T1]{fontenc}    
\usepackage{hyperref}       
\usepackage{url}            
\usepackage{booktabs}       
\usepackage{amsfonts}       
\usepackage{nicefrac}       
\usepackage{microtype}      
\usepackage{lipsum}		
\usepackage{graphicx}
\usepackage{natbib}
\usepackage{doi}
\usepackage{gensymb}
\usepackage{amsmath}
\usepackage{upgreek}
\usepackage{natbib}
\usepackage{amssymb}
\setcitestyle{numbers}

\pdfoutput=1

\newcommand{\Vh}[1]{\ensuremath{\hat{\boldsymbol{ #1}}}}

\newcommand*{\figref}[2][]{\textbf{Fig. \ref{fig:#2}#1}}

\title{Formation of Colloidal Chains and Driven Clusters with Optical Binding.}

\author{ \href{ https://orcid.org/
0000-0002-0829-626X }{\includegraphics[scale=0.06]{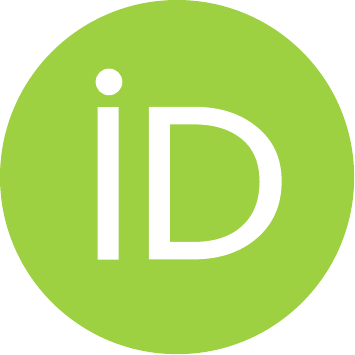}\hspace{1mm}Dominique Davenport}\thanks{Use footnote for providing further
		information about author (webpage, alternative
		address)---\emph{not} for acknowledging funding agencies.} \\
	School of Natural Sciences\\
	University of California - Merced\\
	Merced, CA 95340 \\
	\texttt{ddavenport3@ucmerced.edu} \\
	\And
	\href{ https://orcid.org/
0000-0002-2043-1639 }{\includegraphics[scale=0.06]{orcid.pdf}\hspace{1mm}Dustin Kleckner} \\
	School of Natural Sciences\\
	University of California - Merced\\
	Merced, CA 95340 \\
	\texttt{dkleckner@ucmerced.edu} \\
}

\renewcommand{\shorttitle}{Formation of Colloidal Chains and Driven Clusters with Optical Binding}

\hypersetup{
pdftitle={Formation of Colloidal Chains and Driven Clusters with Optical Binding.},
pdfsubject={physics},
pdfauthor={Dominique Davenport, Dustin Kleckner},
pdfkeywords={},
}

\begin{document}
\maketitle

\begin{abstract}
{We study the effects of the optical binding force on wavelength sized colloidal particles free to move in a counter-propagating beam.  This work is motivated by the concept of using optical binding to direct the assembly of large numbers of colloidal particles; previous work has used small numbers of particles and/or 1D or 2D restricted geometries.  Utilizing a novel experimental scheme, we describe the general static and dynamic self-organization behaviors for 20--100 particles free to move in 3-dimensional space.  We observe the self-organization of the colloids into large optically bound structures along with the formation of driven particle clusters.  Furthermore we show that the structure and behavior of these optically bound systems can be tuned using the refractive index of the particles and properties of the binding light.  In particular, we show that the driven behavior originates from $N$-body interactions, which has significant implications for future work on optically bound clusters of more than 2 particles.}
\end{abstract}


\section{Introduction}

Self-organization is the spontaneous formation of structure via the interactions of their constituent particles \cite{whitesides_self-assembly_2002}.
The resulting structures from a self-organized system can be found in equilibrium and non-equilibrium states; the latter provides a pathway towards active and adaptive matter.
While adaptive materials are common in living systems such as cells, it is possible to artificially introduce dynamic forces which can externally deliver energy to a system while mediating self-organization \cite{grzybowski_self-assembly_2009}.
Dynamic self-organization is relatively new but active field of self-assembly with the goal of extending our knowledge of equilibrium thermodynamics onto living systems.
Some examples include the dynamic assembly of non-biological components including active colloids \cite{ginot_aggregation-fragmentation_2018,wang_adaptive_2015,howse_self-motile_2007}, rotating discs \cite{grzybowski_dynamic_2000,grzybowski_dynamics_2002}, and magnetic swimmers \cite{kokot_dynamic_2017,tierno_controlled_2008,snezhko_magnetic_2011}.

Optical binding is a long-range light-induced force which causes dielectric particles to interact mutually through light-scattering \cite{burns_optical_1989}.
In theory, optical binding can be felt between polarizable molecules ($\sim$ 1 nm) \cite{thirunamachandran_intermolecular_1980} up to biological cells \cite{bai_lateral_2018} ($\sim$ 5 $\upmu$m).
In practice, because the force is relatively weak and requires a strong scattering response, the effects of optical binding are observed most strongly for objects that are on the order of the wavelength of light in size.
Moreover, optical binding can give rise to non-equilibrium forces because the optical field applied to the system is also a constant supply of external momentum which can contribute to particle motion \cite{li_non-hermitian_2021}, although this effect has been largely unexplored in previous experimental work.
In principle optical binding forces can be explained using existing approaches -- forces can be exactly predicted using a multiple scattering calculations -- but nonetheless it remains difficult to predict and explain the behavior of systems composed of many particles  \cite{li_non-hermitian_2021}.

Numerous past studies have illustrated the complexity of the optical binding force.
Two-particle studies have shown that the force can contain multiple stable points defined by multiples of the light wavelength \cite{burns_optical_1989, dapasse_optical_1994,mohanty_optical_2004}.
By increasing the number of particles, previous experimental and computational studies have shown the self-organization of 1-dimensional chains \cite{tatarkova_one-dimensional_2002}, static and drifting 2-dimensional lattices exhibiting stable and quasi-stable behaviors \cite{ng_photonic_2005}, and observations of bistability \cite{metzger_observation_2006} and multistability \cite{karasek_long-range_2008} in the equilibrium positions of optically bound particles \cite{dholakia_colloquium_2010}.
The complexity that arises from optical binding in multi-particle systems can largely be attributed to feedback in the form of interference with the incoming field through multiple scattering effects \cite{dholakia_colloquium_2010}.
One often neglected fact is that optical binding forces are sensitive to $N$-body effects, and can not be treated as strictly pair-wise interactions \cite{rodriguez_optical_2008}.
These $N$-body interactions can lead to highly correlated and emergent behaviors but can consequently be difficult to model.
More recent studies have described complex dynamical assemblies of metallic and dielectric nanoparticles
\cite{forbes_optical_2020, nan_dissipative_2018,brzobohaty_complex_2020}.
These studies suggest that highly nonlinear and dynamic interactions can occur for smaller particles, where interactions are highly pair-wise, because of the complexity of the potential landscapes generated by optical scattering alone \cite{yan_potential_2014}.
The presence of thermal noise in a complex potential landscape can disrupt stability.
Thus, in studying a system which $N$-body forces may be strong \emph{and} potential landscapes complex, we aim to untangle the effects between the two.

In this manuscript we study optical binding of large numbers of particles ($N>$ 20) free to move in three-dimensional space.
We assess which features of the assembly can be described by pairwise interactions verses features which emerge from higher-order effects.
By comparing strong and weak scatterers, we share new insights on the relationship between refractive index, scattering strength, non-conservative forces, and non-linear optical binding.
We present observations of unexpected emergent driven behaviors which appear to manifest from the deviation from pairwise forces.
We also use the coupled dipole method (CDM) \cite{draine_discrete-dipole_1994,hoekstra_radiation_2001}, also known as the discrete dipole approximation (DDA), to explain the origin of these effects.

\section{Experimental Design}
\label{sec:headings}

\begin{figure*}[t!]
    \centering
    \includegraphics{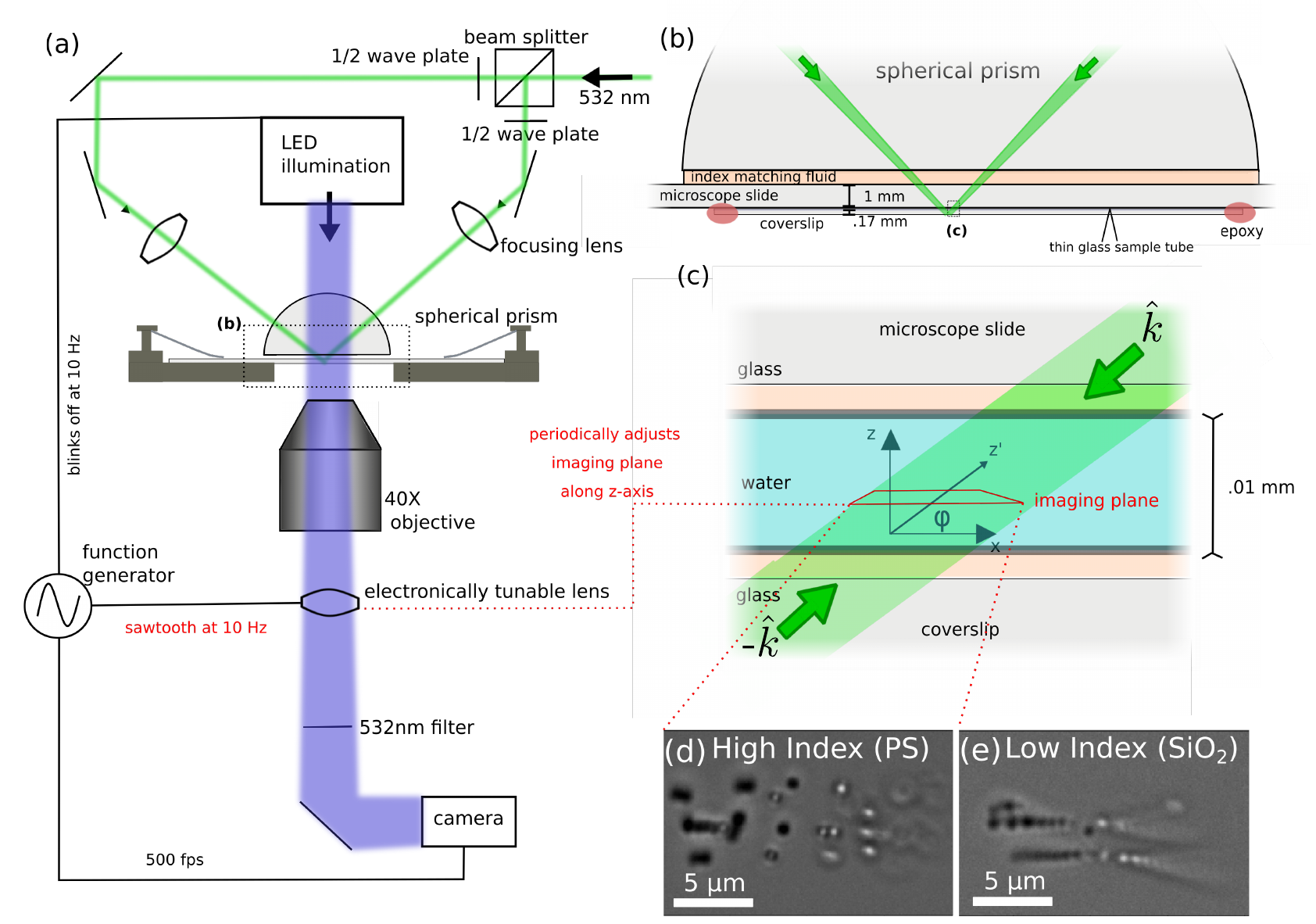}
    \caption{(a) Experimental setup. The laser beam is split into two paths that are aligned to counter propagate through the colloidal sample. Both beams are adjusted so that they are totally internally reflected at the coverslip. The polarization and incident angle of each beam can be adjusted individually. The camera, electronically tunable lens, and illumination source are all synchronized to a wave-function generator for the acquisition of 3-dimensional videos. (b) Depth profile of sample cell, we define z-axis as direction along the sample depth and $\Vh k$ as the axis corresponding to the beam propagation axis. (c) image still of HIPs in the optical field. (d) image still of LIPs in the optical field.}
    \label{fig:opticalpath}
\end{figure*}

\subsection{Experimental parameters}

There are a number of relevant parameters which can alter how an object scatters light and consequentially alter the properties of optical binding interactions.
In practice, the most important quantities are the size parameter, $ka$, and the relative refractive index $m = n_p / n_0$ (where $k$ is the wavenumber of the incident field in the background medium, $a$ is the particle radius, and $n_p/n_0$ is the particle/background medium index of refraction).

Tuning the parameters will not necessarily yield linear responses, which is why we must be careful about generalizing optical binding behavior.
For instance, the size parameter alone can span three different regimes: the Rayleigh limit, $ka\ll$ 1, the ray-optics limit, $ka\gg$ 1, and the Mie scattering regime, $ka \sim$ 1, each of which are separated by drastically different assumptions including the relevance of multiple scattering, scattering modes, and how the strength of the forces scale.
Our current study is focused in the Mie regime.

The relative refractive index, $m$, affects the scattering strength of a single particle, and so modulates the importance of multiple-scattering events \cite{ashkin_observation_1986}.
We use this study as an opportunity to contrast the behaviors of strong and weak scatterers.
We perform a side-by-side analysis of two commonly available colloidal microspheres which differ in refractive index.
We refer to these as high-index particles (HIPs, m=1.2, made of polystyrene) and low-index particles (LIPs, m=1.1, made of SiO$_2$).

Unique to field-driven self-organization, direction and geometry become particularly important.
Because optical scattering is directional, optical binding is an interaction which breaks symmetry along the axis of the field direction.
As a result, experimental and numerical studies have often focused on either two distinct optical binding geometries: lateral \cite{burns_optical_1989, dapasse_optical_1994,mohanty_optical_2004,wei_lateral_2016,ng_photonic_2005,li_non-hermitian_2021} and longitudinal\cite{karasek_analysis_2006,karasek_long-range_2008,karasek_longitudinal_2009,singer_self-organized_2003,tatarkova_one-dimensional_2002} optical binding.
Lateral optical binding describes binding that occurs between particles with a displacement perpendicular to the direction of propagation, $\Vh k$, while longitudinal optical binding describes binding of particles displaced along $\Vh k$.
In the current study, we implement a novel experimental approach to present the self-organization of many particles free to interact in all directions.
We believe this configuration is particularly illuminating for understanding potential bulk behaviors.

Finally, previous studies have focused on how the stability of optically bound matter is affected by damping conditions \cite{ng_photonic_2005,li_non-hermitian_2021} suggesting that the low damping can lead to instabilities due to the systems inability to remove energy.
Because we use sub-micron particles suspended in a fluid, our study is in an over-damped regime.
Our estimated Reynold's number is on the order of $Re = 10^{-6}$ (assuming 500 nm diameter particles in water with a characteristic velocity of 1 $\upmu$m/s).

\begin{figure}[t!]
    \centering
    \includegraphics[scale=.65]{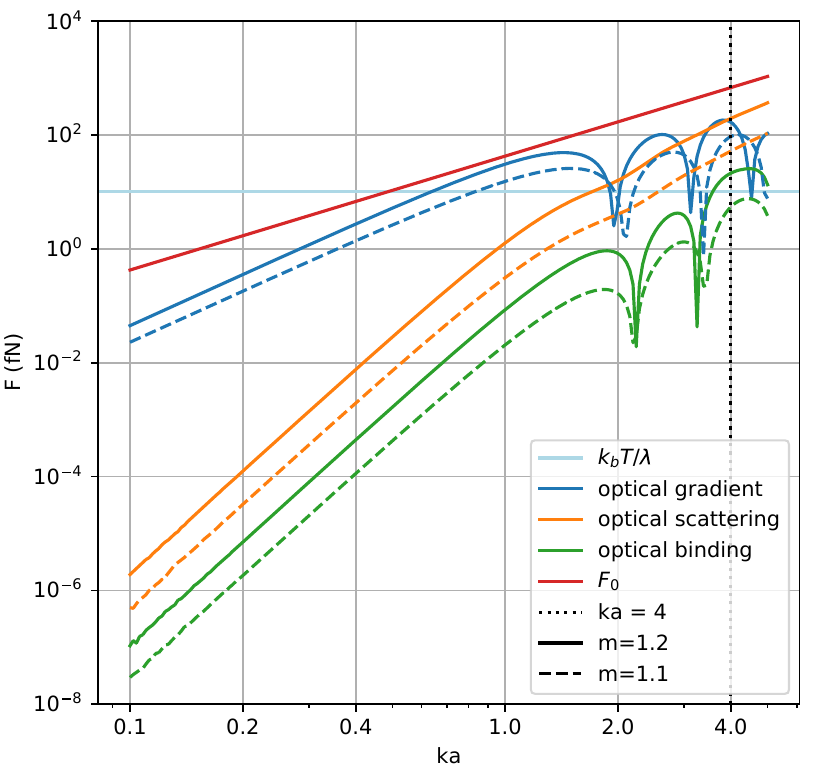}
    \caption{Strength of optical forces for a range of size-parameters ($ka$). The optical gradient force (blue) was numerically simulated by placing a single particle in a node of a standing wave, the optical scattering force (orange) was numerically simulated by placing a single particle in a propagating plane-wave. The optical binding force was obtained by placing to particles laterally in a counter-propagating field at a separation of $2\lambda$. The reference force, $F_0$, is also plotted. In all simulations we assume a light intensity of 1 mW/$\upmu$m$^2$. 
    }
    \label{fig:scale}
\end{figure}

\subsection{Optical setup}

The experiments are performed on a custom designed inverted microscope (\figref{opticalpath} (a)). 
Two Gaussian beams ($\lambda$ = 532 nm in vacuum, or 400 nm in solution) are aligned and focused through the sample tube to generate a counter-propagating crossed polarized optical field ($w_0 $ = 27.0 $\pm$ 0.3 $\upmu $m).
The polarizations of each of the two beams can be adjusted individually which allows us to switch between a counter-propagating standing wave and a crossed polarized configuration.
In this specific setup, a portion of beam power is redirected back towards the laser source which is removed using an optical isolator placed between the laser head and the beam splitter.
The final focusing of the beam allows us to further modulate the beam power density.
The orientation of the beam through the sample is at an angle of approximately $\upphi_g$ = 45\degree{}  through the glass layers and $\upphi$ = 36\degree{} through the water.
We achieve total-internal reflection at the glass-air interface at the bottom surface of the coverslip depicted in \figref{opticalpath} (c).
The sample tube depth is 10 $\pm$ 1 $\upmu$m.
We collect 3-dimensional video microscopy data using a combination of a microscope objective (40X, NA = 0.75) and an electronically tunable lens (ETL).
The scanning of the ETL is synchronized with the imaging camera acquisition to capture 10 volumes per second with 50 frames per volume.
By adjusting the focal length of the ETL, one can adjust the z-depth of the imaging plane shown in \figref{opticalpath} (c).
By driving the ETL approximately 6 $\%$ of its full range, we scan more than the full 10 $\mu m$ of the depth of the sample.
The signal output to the ETL is in the form of a periodic sawtooth pattern (driven at 10 Hz) which increases linearly for 70 $\%$ of the total period and quickly decreases linearly for the remaining 30 $\%$.
As such, the first 70 $\%$ of the period is used to create the full volume.
The 3D scanning method gave us a spatial resolution of 0.8 $\upmu$m/pixel in the z-direction and a time resolution of 10 Hz which was sufficient to reliably locate particles within the volume.

The colloidal particles in the sample exhibit only Brownian motion until they enter the beam region.
Particles that diffuse into the region remain confined within the Gaussian beam but can move thermally in all 3-dimensions within the area of confinement.
Radiation pressure forces, which can be a strong optical force in the direction of the field propagation \cite{ashkin_observation_1986}, are carefully balanced by the two counter-propagating beams.
As nearby colloids drift into the beam, the density of the particles increase slowly.
While there is no direct control to stop diffusive colloids from drifting into the beam, the variability in the number of particles over the entire 2 mins of observation are between 1 and 5 particles ($\sim$ 3--7 \% of total particles).

The light intensity is a parameter which plays a linear role in the strength of the optical binding force and can be tuned quite easily.
For the current study, we adjust the peak light intensity of the Gaussian beam up to $\sim$ 2.5  mW/$\upmu \text{m}^2$.
For the numerical simulations we use a reference intensity of $I_0$ =  1 mW/$\upmu \text{m}^2$.
Subsequently we define a reference force, $F_0$, which is the total optical momentum impinging on a single particle in a plane wave of intensity $I_0$:
\begin{equation}
    F_0 = \frac{I_0 \pi r^2}{c},
\end{equation}
where $r$ is the radius of the particle and $c$ is the speed of light.
The value approximately 650 fN for 500 nm diameter particles.
It is also useful to compare this to thermal excitation.
For an order of magnitude estimate of how the thermal activity compares to the strength of the optical forces in the experiment, we can convert the thermal energy, $k_bT$, to an equivalent force by including a length scale.
Given that the optical binding force has spatial oscillations with the wavelength of the incoming light, the approximate scale of the equivalent thermal force is given by $F_T \sim k_bT / \lambda \sim 10$ fN.

We performed a numerical analysis to predict suitable light intensity and particle parameters which should allow us to observe optical binding effects (\figref{scale}).
By comparing the scaling of optical forces -- including binding -- with an estimated force of thermal excitation, we show that the size parameter is important in determining the strength of optical binding effects relative to thermal motion.
The particles used in this study have a size parameter of $ka = 4$; as can be seen in \figref{scale}, a power density of 1 mW/$\upmu$m$^2$ should ensure that all optical forces are comparable to or larger than the effective thermal forces.

Potentially more important than the strength of the force is the nature of the interaction.
We note that the optical binding force is a combination of second-order gradient and scattering forces.
In \figref{scale}, we plot the relative strength of first-order gradient and scattering forces for HIPs and LIPs to approximate their relative contributions to the binding force.
We show here that the size parameter, $ka$, has a strong effect on the relative balance of scattering and gradient forces, thus we draw a distinction between recent studies done on dielectric nanoparticles \cite{brzobohaty_complex_2020}.
Note that the optical binding force is computed only for a pair of particles spaced by 2$\lambda$; for larger numbers of closely spaced particles it can be 1--2 orders of magnitude stronger.
We observe for $ka = 4$ that the HIPs have a strong scattering force response, relative to the gradient force.
Because scattering forces are generally non-conservative, we expect that it is the stronger presence of scattering forces in optical binding for the HIPs that give them a more dynamic self-organization behavior as opposed to the LIPs.

\subsection{Sample preparation}

Colloidal samples are diluted in water ($\sim$ 0.01 \text{w/v} \%) and placed into thin rectangular Borosilicate tubes (100 {$\upmu$}m wide $\times$ 10 {$\upmu$}m thick $\times$ 50 mm long) purchased from \emph{VitroCom}.
We performed experiments with polystyrene (`HIPs', diameter $d$ = $518 \pm 10$ nm, relative index $m$ = $1.20$ \cite{zhang_complex_2020,hale_optical_1973}) and silicon dioxide (`LIPs', diameter $d$ = $518 \pm 20$ nm, relative index $m$ = $1.10$ \cite{malitson_interspecimen_1965,hale_optical_1973}) purchased from \emph{microParticles GmbH}.
In the experiments, the number of colloidal particles are in the range of 30-50 for HIP experiments and 40-100 for the LIP experiments.
We found that LIPs were more readily collected into the beam, resulting in higher effective particle densities compared to the HIPs even when the initial density is the same.
The tubes allow us to keep a precise sample depth of $10 \pm 1 \upmu$m.
This tube depth is important for confining particles to a range in the z-direction.
The tubes are coated with index matching fluid and placed in-between a coverslip and microscope slide.
Finally, the coverslip is glued to the microscope slide, using UV curable Epoxy (\emph{Norland Optical Adhesive NOA61}), which seals the tube and index matching fluid.
The samples are fixed onto the \emph{Mad City Labs - RM21} microscopy base allowing for micro-precision movement of the sample in the $x$-$y$ plane.\\

\begin{figure*}[t!]
    \centering
    \includegraphics[scale=1]{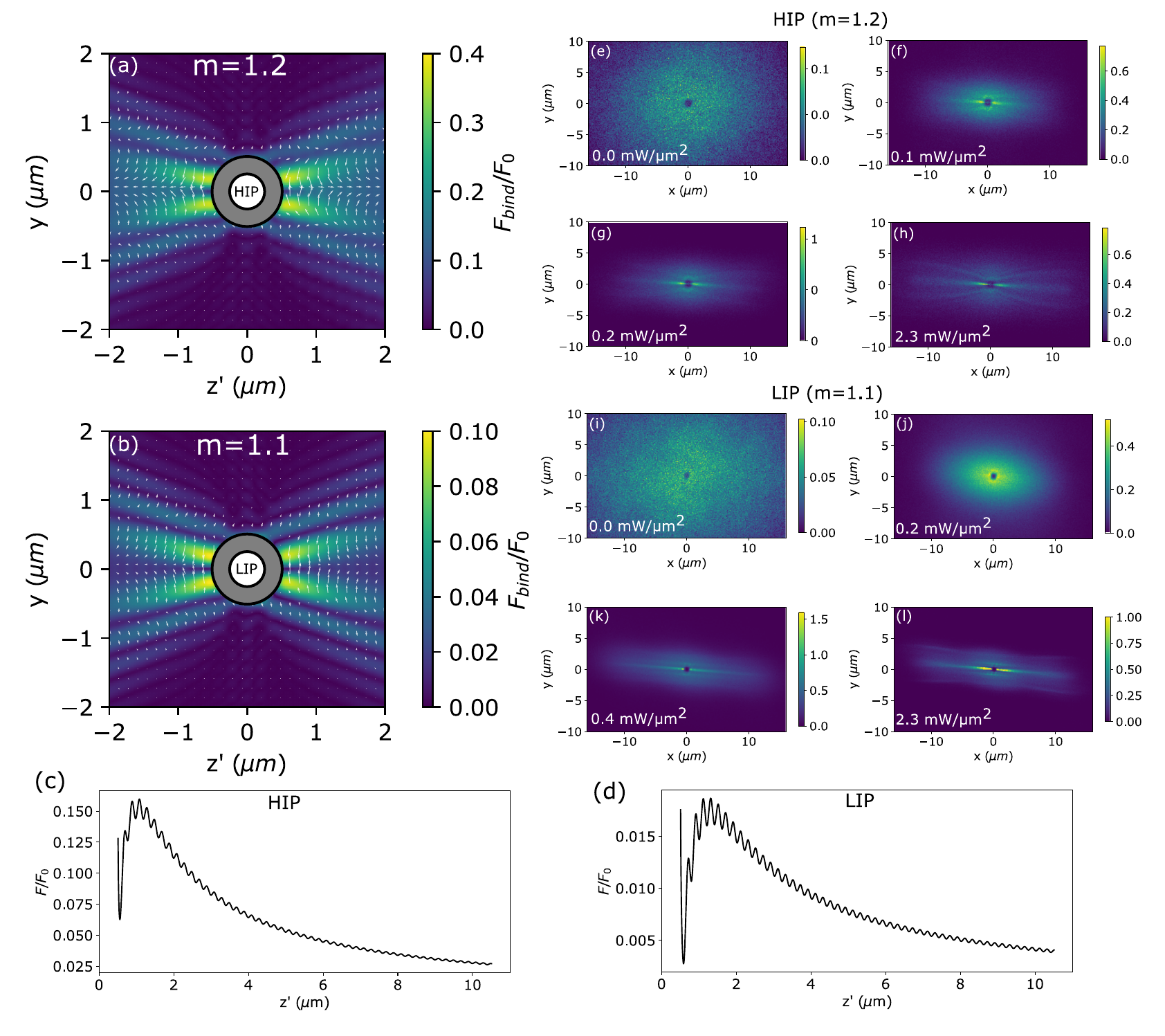}
    \caption{\textbf{(a,b)} Two-body force map of optical binding forces for HIPs and LIPs ($ka = 4$) placed in counter-propagating crossed-polarized plane wave ($\lambda = 0.4 \upmu \text{m}$ in solution) propagating in the $+ \Vh k = +z'$ and $- \Vh k = -z'$ direction.
    Maps are obtained by placing two particles in a CDM simulation and calculating the force applied on each particle at various displacements which fill the map space.
    The forces are provided in a unit-less scale normalized by the reference force, $F_0$, described by the single scattering force on a single particle of the same size and material. The direction of the arrows represent the direction of the force, while the size represents the relative strength.
    \textbf{(c-f)} 2D pair-correlation functions for HIPs in the x-y plane at various light intensities.
    \textbf{(g-j)} 2D pair-correlation functions for LIPs in the x-y plane at various light intensities.
    The color bar in the pair-correlation functions gives the 2D probability density of the particles, and has units of $\upmu$m$^{-2}$.
    }
    \label{fig:static}
\end{figure*}

\section{Results}

\subsection{Observations of self-organization behavior}

Optical binding effects are immediately apparent between particles within the beam region.
For instance, spatial ordering is immediately observable within the beam area -- effects which can be enhanced by increasing the light intensity.
The most apparent feature is the generation of multiple chains of particles aligned along the beam propagation direction (\figref{opticalpath} (c)).
Once collected into a chain, particles within the chain can be observed to move collectively together within the beam region.
The density of particles is large enough to observe the formation of multiple chains at a time.

Other features of the formed structures arise on further examination.
For instance, the HIPs are found to self-organize into extended chains which extend along the sample depth--or--small clusters of typically (3-5) particles long.
Both HIP structures can be observed in \figref{opticalpath} (c) and \dag ESI Movie S1.
Unlike the HIPs, the LIPs tend to form extremely long and close packed chains of particles (rods) (\dag ESI Movie S2).
Two LIP rods can be observed in \figref{opticalpath} (d).
We note that because the chains extend along the entire available depth of the sample, that the sample depth plays a relevant role as a boundary for the system.

\subsection{Comparing observations to two-body forces}

To understand the dominating interactions which are leading to the general behavior of both systems, we numerically generated two-body force maps for the two particle types (\figref{static} (a,b)).
The force maps were created using a CDM based simulation \cite{draine_discrete-dipole_1994,hoekstra_radiation_2001}  (see supplementary material for details).
Comparing force maps for HIP ($m=1.2$) and LIP ($m=1.1$) particles reveals surprisingly little difference, apart from the overall strength.
The strength differences can be observed in \figref{scale} by comparing how the optical forces are generally stronger for HIPs.
The strength can be found to scale approximately like the reflectivity of a dielectric plane, which scales like $\sim(m - 1)^2$.

Despite this, the two-body force maps can be used to describe some of the overall structures observed in the system.
The long range alignment along the beam propagation for the HIPs and LIPs are qualitatively consistent with the two-body force maps (\figref{static}), which suggest strong forces can tend to pull particles onto the axis of propagation ($\Vh k$).
The $\Vh k$ axis, as shown in \figref{opticalpath} (b) is rotated 54\degree{} from $z$.
For comparison, we approximate the energy $F_0 \lambda \sim 65 k_bT$, thus the HIPs can experience $ \sim 20 k_bT$ of energy keeping them aligned in the $\Vh k$ axis in the beam at 1 mW/$\upmu$m$^2$.
The LIPs experience $ \sim 5 k_bT$ at 1 mW/$\upmu$m$^2$.

The HIPs force-map shows a repulsive force between particles that are aligned along the field propagation (\figref{static} (a)).
This fits with the observation of HIPs forming extended particle chains (\figref{opticalpath} (c)), rather than tightly packing.
What the force-map fails to predict for the HIPs are the presence of the smaller tightly packed HIP clusters also found in the experiment.
While it is expected that we observe structures that cannot be predicted from two-body optical binding interactions, here we observe an emergence of a force which is not only strongly contributing, but can act in the opposite direction of the two-body force.
For the HIPs, the two-body force is repulsive along the \Vh{k} axis but we observe experimentally an unexpected close-range attractive force suddenly drives particles together into small clusters.

\subsection{Pair-correlation function}

To map the time-averaged structure created by optical binding forces, we compute a 2D pair correlation function (2D PCF).
Given all the 2D particle positions, $\vec r_i$, we can compute all relative displacements $\vec \Delta_{ij} = r_i - r_{j \neq i}$; the 2D pair correlation function is then the histogram of these displacements, averaged over all frames in the data set and normalized by the bin size (so that the result is expressed as a 2D density).
The form of the pair-correlation function also gives an approximation for an effective potential through which the particles interact.
We collected the data for the pair-correlation function by recording the positions of colloidal particles all located within the same weakly focused beam of light.
The beam is fixed at a constant light intensity for a duration of 2 mins during which we collected 3-dimensional imaging data.
3-dimensional volumes were used to locate particles over the entire 10 $\upmu$m in depth.
However, as the optical configuration results in poor resolution in z, we summed over the z-axis to generate the 2-dimensional plots.
As such, the \Vh{k} axis is projected along the x-axis.
The particles were located using the \emph{trackpy} implementation of Crocker and Grier \cite{crocker_methods_1996, allan_trackpy_2015}.
While the particle speeds and dense particle clustering made it challenging to track particles through time, we found that the algorithms were quite efficient at locating particles within a frame.
We found relatively low fluctuations in the total number of particles from frame to frame suggesting that the particle locating algorithms were performing consistently.
Inter-particle displacements were determined by particle locations given for each time-step, thus the frequencies of each displacement over the entire run were available.

The 2D pair-correlation function (PCF) for the HIPs suggest an increase in spatial order with the increase in light intensity \figref{static} (e-h).
The form of the PCF shows the dominating feature that the HIPs tend to align along the beam propagation.
At higher light intensities, $I >$ 1 mW/$\upmu$m$^2$, multiple lines off-axis begin to appear.
This not only suggests that particles are interacting to form long range structures, but that there are optical binding interactions occurring in multiple directions within the field.
For example, at high powers there is clear evidence of preferred inter-chain transverse spacing of $\sim 1\ \upmu$m.
This can be explained by the computed pairwise force diagrams, which have a converging force in the y direction at these separations.
The two-body force maps (\figref{static} (a-b)) can be used to help explain the multiple off-axis lines that appear in the PCF at higher light intensities, as we can observe multiple off-axis lines in which the force arrows converge.

The 2D PCF for the LIPs are very consistent with the observation that the particles collapse into tightly bound rods (\figref{static} (i-l)).
One feature that becomes prominent are secondary off-axis lines at higher light intensities.
The secondary lines are evidence of multiple rods interacting to form a long-range regular spacing over time.

\begin{figure}[t!]
    \centering
    \includegraphics[scale=1.2]{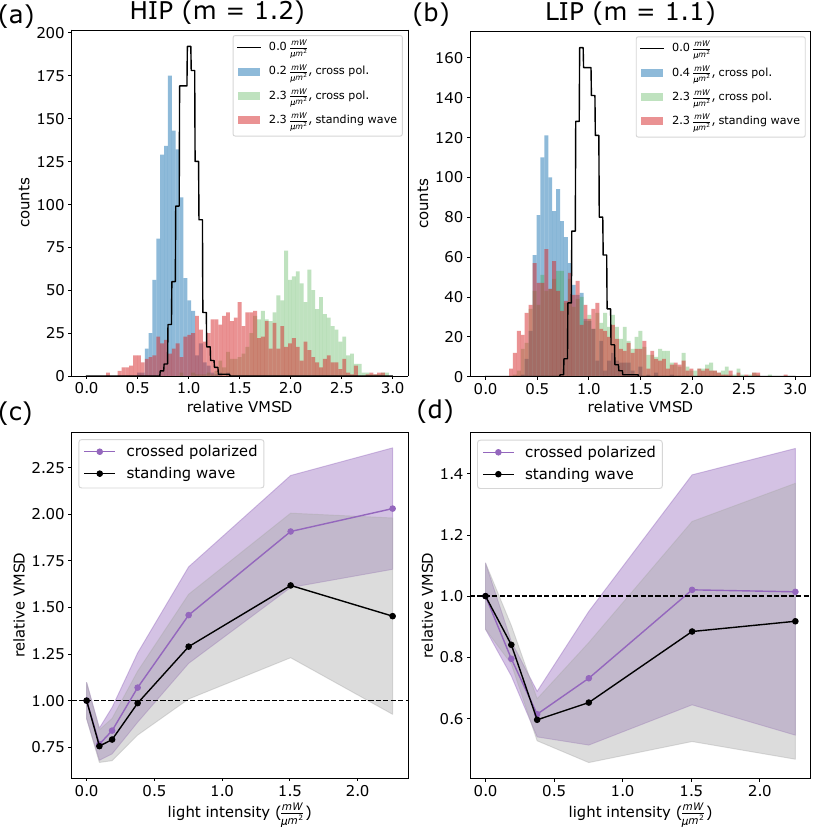}
    \caption{\textbf{(a)} Distribution of relative FMSD values over 2 min duration for various intensities. For the high light intensity case (2.3 mW/$\upmu$m$^2$) we share a comparative distribution of FMSD values for the same system in a counter-propagating standing wave. \textbf{(b)} The average relative FMSD value over 2 minutes for various light intensities. The shaded regions represent the variability in the relative FMSD values over the 2 minutes. Values over the dotted line represent the average motion of the particles greater than what would be observed in a purely diffusive system. \textbf{(g)} Distribution of relative FMSD values over 2 min duration for various light intensities. For the higher light intensity case (2.3 mW/$\upmu$m$^2$) we share a comparative distribution of FMSD values for the same system in a counter-propagating standing wave.  \textbf{(h)} The average relative FMSD value over 2 minutes for various light intensities. The shaded regions represent the variability in the relative FMSD values over the 2 minutes. Values over the dotted line represent the average motion of the particles greater than what would be observed in a purely diffusive system.
    }
    \label{fig:dynamic}
\end{figure}

\subsection{Volume Mean Squared Difference (VMSD)}

\begin{figure}[t!]
    \centering
    \includegraphics[scale=0.8]{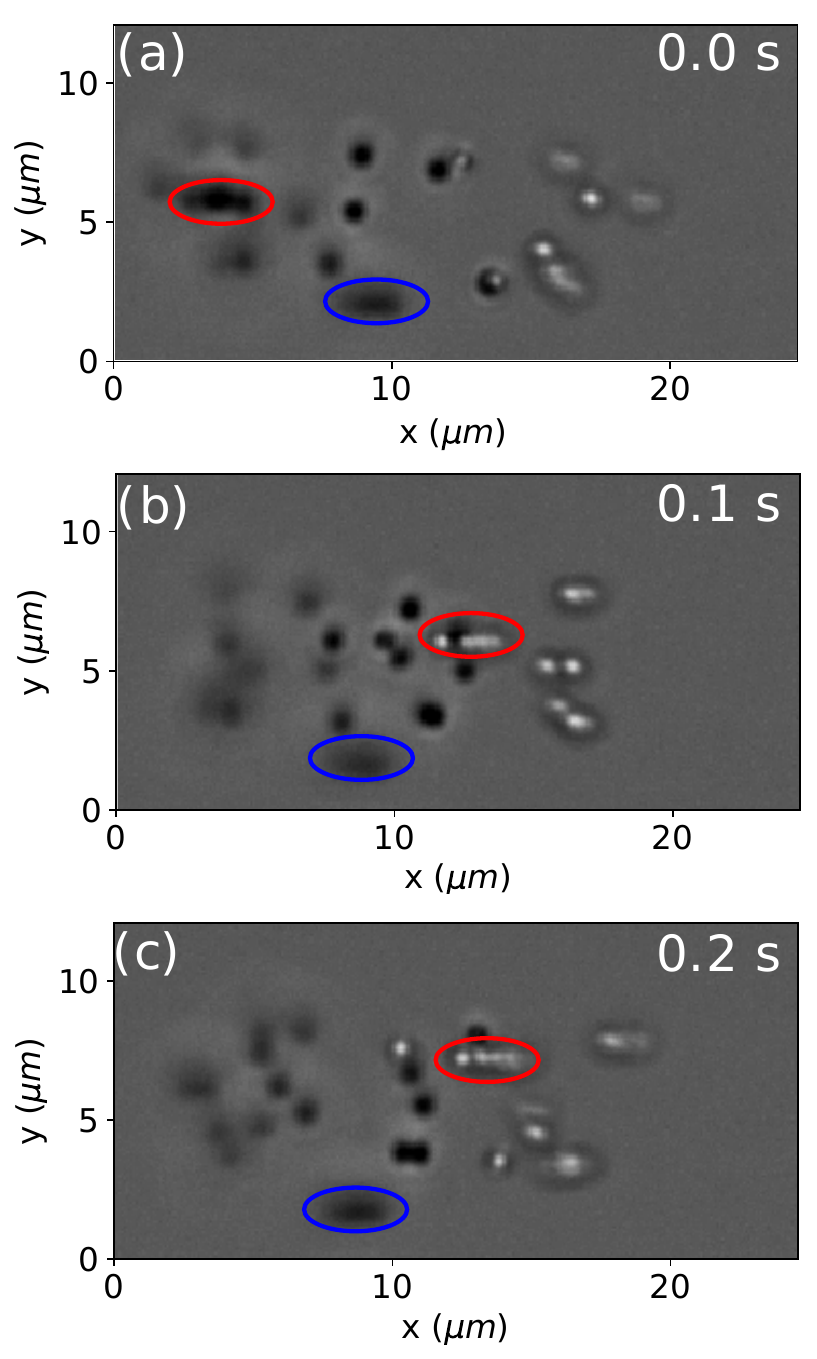}
    \caption{\textbf{(a-c)} Three snapshots of the optically bound HIPs over a 0.2 s duration. Two clusters are circled: (red) cluster of 5 particles ballistically moving (>80 $\upmu \text{m/s}$) through the sample and (blue) isolated cluster of 4 particles remaining relatively still.}
    \label{fig:img_stills}
\end{figure}

We use the mean squared displacement of particles to determine their dynamic behavior.
Due to high-density particle clustering and the high velocities of clusters evidenced in \figref{img_stills}, it is difficult to reliably track individual particles through time.
This is especially true for close-packed clusters, which -- as we shall show later -- appear to drive motion in the HIPs.
Instead of attempting to directly track the particles in time, we use a proxy for the mean squared displacement by obtaining the mean squared difference between two subsequent image volumes.
We found that the volume mean squared difference (VMSD) between subsequent frames gives us more consistent measures of motion while remaining highly correlated with the true mean squared displacement (details of the method are described in the supplementary material).
The VMSD, $\Delta$, is given by:
\begin{equation}
    \Delta(t) = \frac{1}{N_p(t)} \left[\sum_{i,j,k}^{volume}(p_{i,j,k,t} - p_{i,j,k,t-dt})^2  - \Delta_{0}\right]
\end{equation}
where $p_{i,j,k,t}$ is the pixel value at a given location and time, $dt = 0.1$ is the time between frames, and $\Delta_0$ is the background VMSD caused primarily by camera noise, and is computed for when there are no moving particles in the frame.

As shown in the supplemental materials, $\Delta$ is proportional to the squared particle displacement provided the particles average motion between frames is smaller to or comparable the particle size.
In practice, we normalize this value relative to one obtained by turning the optical binding laser off, in which case the particles experience only Brownian motion.

The distribution average particle motion for the HIPs is provided in \figref{dynamic} (a).
At lower light intensities, we find narrow distributions of the average particle motion and average values lower than what is found in a Brownian system, reflecting the fact that the particles are being confined by the optical binding forces.
At higher light intensities, we find that the distributions shift to higher average values and the size of the fluctuations are greater, indicating the presence of a non-conservative driving force.

Surprisingly, comparing the pair-correlation function to the VMSD values suggests that the average motion is increasing even as the particles are becoming more ordered.
This would not be expected for a conservative pair-wise force; in this case ordering will result in weaker fluctuations.
We do indeed observe this for lower power levels (<0.4 mW/$\upmu$m$^2$).
Above these power levels, the increasing motion suggests that we are forming collections of particles which experience additional non-conservative forces from the optical field.

We found that the behavior can be altered by aligning the polarizations of the counter-propagating beams to generate a standing wave pattern.
In this configuration, the optical binding area includes multiple planes of high light intensity perpendicular the the propagation axis and separated by $\lambda/2$.
While the overall average motion is suppressed in the standing wave, the system is still observed to fluctuate strongly between low average motion ($\sim$ 0.5 $\times$ Brownian motion) and high average motion ($\sim$ 2.5 $\times$ Brownian motion).
At the same light intensities to the previous configuration, the standing wave had the effect of dramatically suppressing the overall motion (\dag ESI Movie 3).

Using the same range of intensities, the same analysis was done for the LIP system (\figref{dynamic} (b,d)).
At lower light intensities, we find narrow distributions of the average particle motion and mean values lower than what is found in a Brownian system.
At higher light intensities, the average values do not exceed what is found in the Brownian system; however the distributions are far less Gaussian.
The distribution suggests that the average motion is low with occasional rare events that lead to high motion.
The LIPs in the standing wave do not show significant differences than particles in the cross-polarized counter-propagating beam.

Comparing the dynamic behavior between the HIPs and LIPs, there are clear differences between the two systems.
The optical binding force when increased is shown to significantly increase the average kinetic energy of the system of HIPs.
This implies that optical binding is a source of non-conservative motion that is especially present in the HIP case.
We believe that non-conservative second-order scattering forces, expected to be stronger for the HIPs, is contributing to the higher overall motion.
The LIPS are instead dominated by second-order gradient forces thus pack into the rod structures until density limitations require the formation of new rods.
Where the standing-wave generates first-order gradient forces which compete with the second-order scattering forces to suppress the overall motion, the same standing-wave does not show much of an effect on the LIPs which are already dominated by gradient forces.

\subsection{Pair-correlation function difference}

\begin{figure*}[t!]
    \centering
    \includegraphics{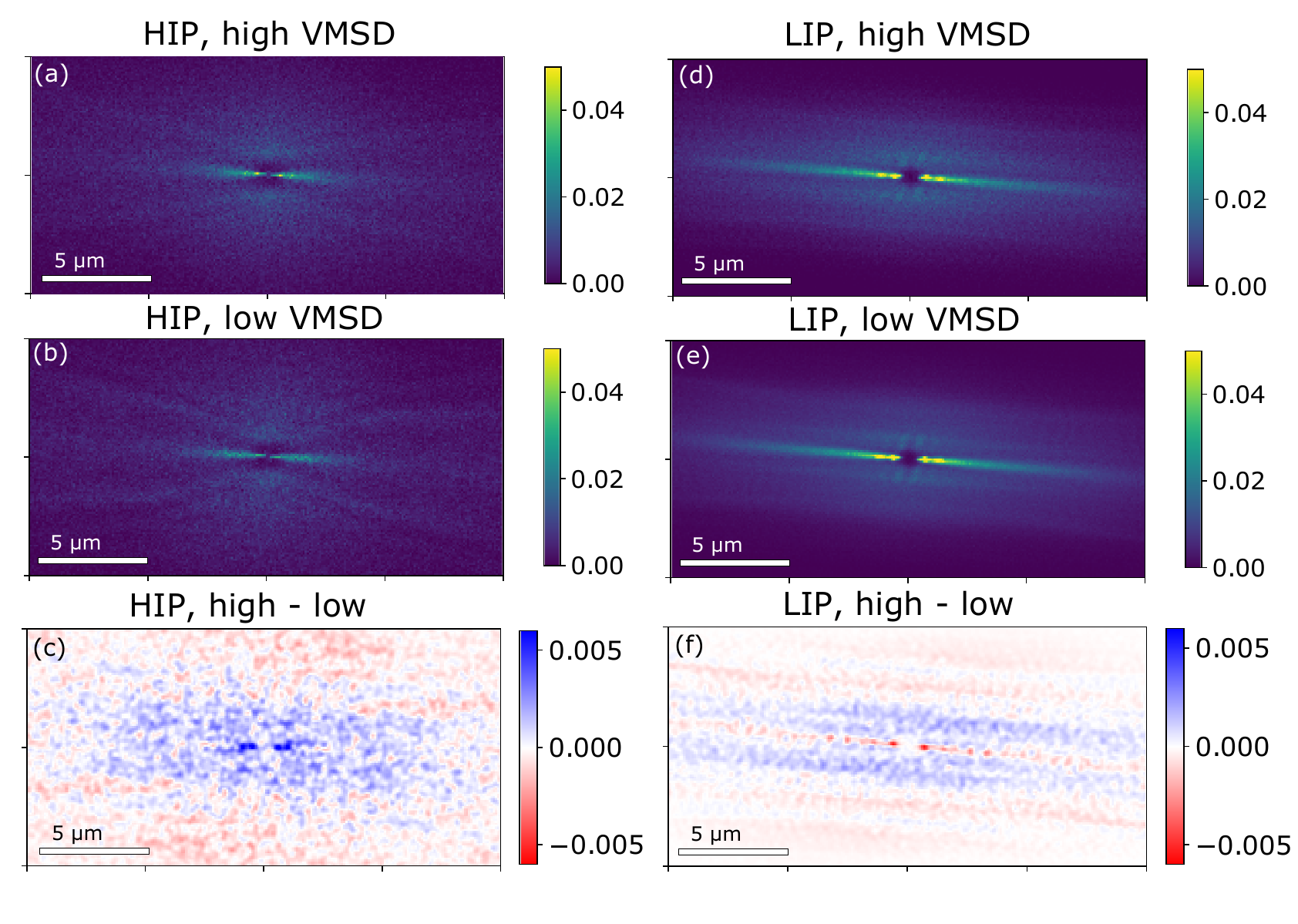}
    \caption{\textbf{(a)} HIP PCF for frames with higher (top 30\%) FMSD values. \textbf{(b)} HIP PCF for frames with lower (bottom 30\%) FMSD values. \textbf{(c)} The difference between the high FMSD and low FMSD pair-correlation functions for HIPs. \textbf{(d)} LIP PCF for frames with higher (top 30\%) FMSD values. \textbf{(e)} LIP PCF for frames with lower (bottom 30\%) FMSD values. \textbf{(f)} The difference between the high FMSD and low FMSD pair-correlation functions for LIPs. Displacements found more often in the higher FMSD frames are positive (blue) while displacements found more often in the lower FMSD frames are negative (red). The PCFs are normalized by dividing the 2D particle density by the number of particles.}
    \label{fig:PCF}
\end{figure*}

We see clear dynamic differences between the HIPs and LIPs; in particular, the HIPs are observed to experience higher average kinetic motion at high powers and a wider range of fluctuations of that behavior.
In each case, we can use the fact that the kinetic motion does fluctuate to subset the data into high motion frames and low motion frames.
Doing so gives us a method to compare particle configurations which may be correlated with driving instability in the system.
To do so, we sort frames from a single video by the FMSD value which represents the average kinetic motion.
We can use each subset of frames to generate PCFs for the top 30\% and the bottom 30\% FMSD values.
Because of the wide distribution of FMSD values, we focus on particles in a high powered (2.3 mW/$\upmu$m$^2$) standing wave.

Comparing the PCFs for the high FMSD frames (\figref{PCF} (a)) versus the low FMSD frames (\figref{PCF} (b)) for the HIPs shows subtle differences.
Primarily, we find more defined features for the low FMSD PCF.
This suggest that the system fluctuates between slower moving ordered states to faster moving disordered states.
Indeed, we observe sudden collapses and regeneration into order states in these optically bound systems.

Secondly, we observe that there is a higher distribution of particles in close proximity for the faster moving system.
We illuminate this subtle difference by subtracting the low FMSD PCF from the high FMSD PCF shown in \figref{PCF} (c).
We not only find that that the overall system moves faster when HIPs are generally closer together, we also find that the correlation is strongest for small clusters. 
We propose that multiple scattering effects which are stronger when particles are at close-range could be driving the collapse of stability and increase in average kinetic motion of this system.

For the LIPs, we find that the high FMSD PCF (\figref{PCF} (d)) versus the low FMSD PCF (\figref{PCF} (e)) are both structurally similar.
Not until taking the difference between the two as shown in \figref{PCF}(f), do we see that there are subtle differences.
For the low FMSD PCF, we observe that particles tend to be distributed among multiple lines: a single line that passes the origin and multiple off-axis lines.
These lines are evidence of the self-organization into multiple rods.
The center line represent displacements between particles that belong to the same rod and the off-axis lines represent displacements between particles that are located in neighboring rods.
Interestingly, the differences between the high and low PCFs suggest that inter-rod distance may play a role in the average kinetic motion of this system.
For instance, one can observe in \figref{PCF}(f) that the the first off-axis line is closer to the center line for the higher FMSD frames. 

Comparing the HIPs and the LIPs is useful for understanding how multiple scattering is effecting these systems at different scales.
For the HIPs, where scattering is much stronger, we find that the presence of small clusters are generally correlated with higher kinetic motion and less structural order of the overall system.
We propose that HIP clusters can be treated as an emergent species which can alter the system dynamics.
On the other hand, the LIPs can generate much larger stable structures before the system fluctuates dynamically. 

\begin{figure*}[t!]
    \centering
    \includegraphics[height=8.5cm,width=16cm]{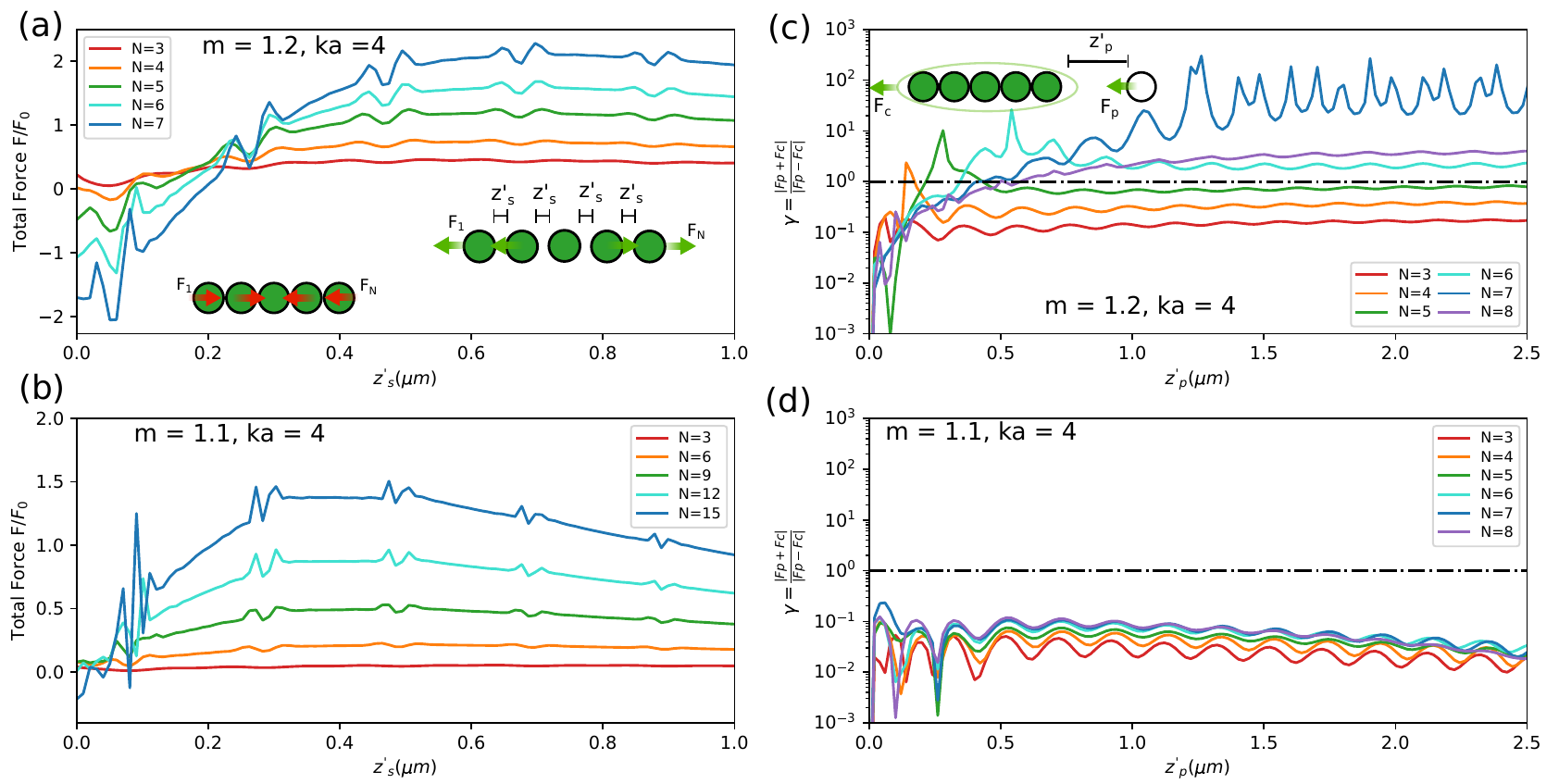}
    \caption{ (a,b) 1-dimensional simulation of forces on particle chains. Particles are placed into a equidistant configuration aligned with the beam axis of propagation (\Vh{k} = \Vh{z}') with inter-particle distance, $z_s'$. At each inter-particle spacing, the total outward force is calculated. A positive force represents a net-force pushing particles away from each other while a negative force represents a net-force in which particles are attracted. We performed the simulation for (a) HIPs ($m$=1.2, $ka$ = 4) and (b) LIPs ($m$=1.1, $ka$ = 4) for various cluster sizes (N) (c,d) 1-dimensional simulation of forces between clusters and neighboring particles. A cluster of N touching spheres is aligned along the axis of propagation with a single particle and the absolute sum over the absolute difference of force on the cluster and the force on the particle are plotted for various particle-cluster spacings, $z'_p$. Simulations were performed for (c) HIP ($m=1.2$, $ka = 4$) and (d) LIP ($m=1.1$, $ka = 4$) particles.}
    \label{fig:numerical}
    
\end{figure*}

We see clear dynamic differences between the HIPs and LIPs; in particular, the HIPs are observed to experience higher average kinetic motion at high powers and a wider range of fluctuations of that behavior.
In each case, we can use the fact that the kinetic motion does fluctuate to subset the data into high motion frames and low motion frames.
Doing so gives us a method to compare particle configurations which may be correlated with driving instability in the system.
To do so, we sort frames from a single video by the VMSD value which represents the average kinetic motion.
We can use each subset of frames to generate PCFs for the top 30\% and the bottom 30\% VMSD values.
Because of the wide distribution of VMSD values, we focus on particles in a high powered (2.3 mW/$\upmu$m$^2$) standing wave.

Comparing the PCFs for the high VMSD frames (\figref{PCF} (a)) versus the low VMSD frames (\figref{PCF} (b)) for the HIPs shows subtle differences.
Primarily, we find more defined features for the low VMSD PCF.
This suggest that the system fluctuates between slower moving ordered states to faster moving disordered states.
Indeed, we observe sudden collapses and regeneration into order states in these optically bound systems.

Secondly, we observe that there is a higher distribution of particles in close proximity for the faster moving system.
We illuminate this subtle difference by subtracting the low VMSD PCF from the high VMSD PCF shown in \figref{PCF} (c).
We propose that multiple scattering effects which are stronger when particles are at close-range could be driving the collapse of stability and increase in average kinetic motion of this system.

For the LIPs, we find that the high VMSD PCF (\figref{PCF} (d)) versus the low VMSD PCF (\figref{PCF} (e)) are both structurally similar.
Not until taking the difference between the two as shown in \figref{PCF} (f), do we see that there are subtle differences.
For the low VMSD PCF, we observe that particles tend to be distributed among multiple lines: a single line that passes the origin and multiple off-axis lines.
These lines are evidence of the self-organization into multiple rods.
The center line represent displacements between particles that belong to the same rod and the off-axis lines represent displacements between particles that are located in neighboring rods.
Interestingly, the differences between the high and low PCFs suggest that inter-rod distance may play a role in the average kinetic motion of this system.
For instance, one can observe in \figref{PCF} (f) that the the first off-axis line is closer to the center line for the higher VMSD frames. 

Comparing the HIPs and the LIPs is useful for understanding how multiple scattering is effecting these systems at different scales.
For the HIPs, where scattering is much stronger, we find that the presence of small clusters are generally correlated with higher kinetic motion and less structural order of the overall system.
We propose that HIP clusters can be treated as an emergent species which can alter the system dynamics.
On the other hand, the LIPs can generate much larger stable structures before the system fluctuates dynamically. 

\subsection{Pathway to cluster formation}

The two-body force maps cannot describe how a cluster or rod forms; in fact, the two-body force maps suggest a weak repulsive force among particles that are aligned along the beam propagation.
As a result, we believe the cluster formation can only be explained by considering a complete $N$-body force which we compute using the CDM.

We find numerically that when the separation between aligned particles is sufficiently large ($\gtrsim 0.25 \upmu$m) the total force is always repulsive for HIP chains and increases in strength with the number of particles (\figref{numerical} (a)).
This suggests that a chain of particles aligned along the propagation axis should remain spread apart.
This is consistent with some of the structures found in the experiment: the lower density HIP chains are often found extending from the lower to upper boundary of the sample.
The strong repulsion at long range could also act as a barrier to cluster formation at very high light intensity and low particle density.
The trend is similar for LIP chains $\upmu$m (\figref{numerical} (b)).

At smaller distances, we observe a decrease in repulsive strength for all HIP chains.
For HIP chains of more than 3 particles, we observe an overall change in sign of the force (\figref{numerical} (a)).
This result suggests that the two-body repulsive force (\figref{static} (a)) dominates as long as the particles are far away; however, when many HIPs become in close range, $N$-body forces are strong enough to overcome two-body repulsion and switch the sign of the force entirely.

The same analysis of the LIP chains of particles shows a similar reduction of the repulsive force at shorter distances; however, an attractive force does not appear until the number of particles exceeds $N=14$ (\figref{numerical} (b)).
The initial reduction of the force, even at small $N$, is indicative of the presence of non-pairwise forces; however the effects are clearly weaker in comparison to the HIPs.

The numerical results provide a potential pathway to the previously unexpected formation of clusters.
The results also highlight the unique nature of the $N$-body optical binding forces.
These effects are shown to be comparatively much stronger for the HIPs.
This supports the hypothesis that a major factor driving the differences in behavior between the HIP and LIP systems is a stronger presence of $N$-body forces due to the higher scattering strength of HIPs.

\subsection{Non-conservative forces on clusters}

We simulated the effects of a particle cluster as it grows in size and interacts with neighboring particles.
We simulate a simplified configuration of N-particle closed packed cluster in line with a single lone particle at varying distance.
We then compute the total force on the clustered particles, $F_c$, as well as the force on the lone particle, $F_p$.
To compare the relative presence of conservative and non-conservative forces, we use the following quantity:
\begin{equation}
    \gamma = \frac{|F_c + F_p|}{|F_c-F_p|}
\end{equation}
If the forces are equal and opposite -- indicating a conservative interaction -- we will obtain $\gamma = 0$.
In contrast, if the system is equal but in the same direction, this would be indicative of a system which would be driven ($\gamma \rightarrow \infty$).

Observing $\gamma$ for the interaction between the HIP cluster and particle, we find that the value increases exponentially as the cluster size increases from 2-6 particles (\figref{numerical} (c)).
At $\gamma > 1$ the force between the cluster and the particle are in the same direction, while for $\gamma \gg 1$ the forces are highly non-conservative.

For HIPs, there are configurations which could lead to uni-directional interactions first appearing for clusters of size 4.
For clusters of sizes $> 5$ particles, $\gamma > 1$ at long ranges.
This result is in good agreement with our observations of driven HIP clusters.
At certain cluster sizes, non-conservative forces appear to allow clusters of particles to propel in an otherwise symmetric initial field.
Despite the particles being identical in the simulation, the cluster-particle interaction adds an asymmetry which can drive the system into unidirectional motion in some cases.
Here we study a limited set of configurations due to their common observable presence in the experimental data; however, we expect there are a number of alternative configurations that can also exhibit highly propelled motion.
We expect that these strongly non-conservative interactions are cause for the increased average motion in the high light intensity binding experiments (\figref{dynamic} (a,b)).

For a similar set of cluster sizes in the LIP case, we find $\gamma < 0.1$ regardless of cluster size (\figref{numerical} (d)).
Indeed, we observed in the experiments that LIPs can build themselves into extremely long rods without exhibiting strongly non-conservative responses at small cluster sizes, as we found for the HIPs.
It is likely that non-conservative forces won't have a strong presence for LIPs until they've assembled into elongated rods for which we do observe fluctuations in the VMSD values.

\section{Conclusion}

Optical binding of many particles in 3D gives rise to complex behavior that can not be predicted from two-body interactions alone.
A simplified view of the interaction does, however, predict some features of the resulting structures: in particular the formation of extended chains of particles along the axis of light propagation.
However, comparing low to high index particles -- whose two-body interaction is nearly identical apart from overall strength -- already demonstrates differences that must arise from $N$-body interactions.
Our results stress how the relative strength of gradient and scattering forces play a role in these systems.
For instance, $ka \ll 1$ particles have been shown to experience interesting conservative interactions on complex potential landscapes; however exploring larger $ka$ affords the capability of controlling non-negligible contributions of non-conservative binding forces.
It is at this size regime that we are able to find considerable differences between the LIPs and HIPs.
In particular, the lower index particles form close packed (or nearly closed packed) chains, while the higher index particles are aligned with the beam, but usually spaced out by several particle diameters.
More strikingly, the dynamic behavior of these particles demonstrates effects that can only be explained by considering non-conservative $N$-body interactions.
This is most dramatically demonstrated by the temporary formation of close-packed clusters of 4-5 high index particles, which appear to drive an instability of the system as a whole.

Our results suggest that although optical binding can be used to guide the self-organization of colloidal particles, the effects of $N$-body forces are critical for the assembly of many-particle systems
when their scattering strength is sufficiently strong.
Notably, this is distinct from nano-particle assembly with optical binding, in which pairwise forces are generally a good approximation \cite{forbes_optical_2020}.
Potentially, this more complex force landscape offer new possibilities: with proper system design these forces could be used to promote or evade specific configurations or produce driven arrangements with dynamic behavior.
Unfortunately, predicting the behavior of large scale optically bound systems remains a difficult task without a quick and accurate method of computing the forces for a fully coupled $N$-body system.
Efforts on this front can be aided by further experimentation which can validate assumptions regarding $N$-body and non-conservative effects over a larger parameter space than the current study.

\bibliographystyle{unsrtnat}
\bibliography{ms}  

\begin{thebibliography}{38}
\providecommand{\natexlab}[1]{#1}
\providecommand{\url}[1]{\texttt{#1}}
\expandafter\ifx\csname urlstyle\endcsname\relax
  \providecommand{\doi}[1]{doi: #1}\else
  \providecommand{\doi}{doi: \begingroup \urlstyle{rm}\Url}\fi

\bibitem[Whitesides and Grzybowski(2002)]{whitesides_self-assembly_2002}
George~M. Whitesides and Bartosz Grzybowski.
\newblock Self-{Assembly} at {All} {Scales}.
\newblock \emph{Science}, 295\penalty0 (5564):\penalty0 2418--2421, March 2002.
\newblock ISSN 0036-8075, 1095-9203.
\newblock \doi{10.1126/science.1070821}.
\newblock URL \url{https://www.science.org/doi/10.1126/science.1070821}.

\bibitem[Grzybowski et~al.(2009)Grzybowski, Wilmer, Kim, Browne, and
  Bishop]{grzybowski_self-assembly_2009}
Bartosz~A. Grzybowski, Christopher~E. Wilmer, Jiwon Kim, Kevin~P. Browne, and
  Kyle J.~M. Bishop.
\newblock Self-assembly: from crystals to cells.
\newblock \emph{Soft Matter}, 5\penalty0 (6):\penalty0 1110, 2009.
\newblock ISSN 1744-683X, 1744-6848.
\newblock \doi{10.1039/b819321p}.
\newblock URL \url{http://xlink.rsc.org/?DOI=b819321p}.

\bibitem[Ginot et~al.(2018)Ginot, Theurkauff, Detcheverry, Ybert, and
  Cottin-Bizonne]{ginot_aggregation-fragmentation_2018}
F.~Ginot, I.~Theurkauff, F.~Detcheverry, C.~Ybert, and C.~Cottin-Bizonne.
\newblock Aggregation-fragmentation and individual dynamics of active clusters.
\newblock \emph{Nature Communications}, 9\penalty0 (1):\penalty0 696, February
  2018.
\newblock ISSN 2041-1723.
\newblock \doi{10.1038/s41467-017-02625-7}.
\newblock URL \url{https://www.nature.com/articles/s41467-017-02625-7}.

\bibitem[Wang et~al.(2015)Wang, Shi, Huang, and Yan]{wang_adaptive_2015}
Andong Wang, Wenyue Shi, Jianbin Huang, and Yun Yan.
\newblock Adaptive soft molecular self-assemblies.
\newblock \emph{Soft Matter}, 12\penalty0 (2):\penalty0 337--357, December
  2015.
\newblock ISSN 1744-6848.
\newblock \doi{10.1039/C5SM02397A}.
\newblock URL
  \url{https://pubs.rsc.org/en/content/articlelanding/2016/sm/c5sm02397a}.

\bibitem[Howse et~al.(2007)Howse, Jones, Ryan, Gough, Vafabakhsh, and
  Golestanian]{howse_self-motile_2007}
Jonathan~R. Howse, Richard A.~L. Jones, Anthony~J. Ryan, Tim Gough, Reza
  Vafabakhsh, and Ramin Golestanian.
\newblock Self-{Motile} {Colloidal} {Particles}: {From} {Directed} {Propulsion}
  to {Random} {Walk}.
\newblock \emph{Physical Review Letters}, 99\penalty0 (4):\penalty0 048102,
  July 2007.
\newblock \doi{10.1103/PhysRevLett.99.048102}.
\newblock URL \url{https://link.aps.org/doi/10.1103/PhysRevLett.99.048102}.

\bibitem[Grzybowski et~al.(2000)Grzybowski, Stone, and
  Whitesides]{grzybowski_dynamic_2000}
Bartosz~A. Grzybowski, Howard~A. Stone, and George~M. Whitesides.
\newblock Dynamic self-assembly of magnetized, millimetre-sized objects
  rotating at a liquid–air interface.
\newblock \emph{Nature}, 405\penalty0 (6790):\penalty0 1033--1036, June 2000.
\newblock ISSN 1476-4687.
\newblock \doi{10.1038/35016528}.
\newblock URL \url{https://www.nature.com/articles/35016528}.

\bibitem[Grzybowski et~al.(2002)Grzybowski, Stone, and
  Whitesides]{grzybowski_dynamics_2002}
Bartosz~A. Grzybowski, Howard~A. Stone, and George~M. Whitesides.
\newblock Dynamics of self assembly of magnetized disks rotating at the
  liquid–air interface.
\newblock \emph{Proceedings of the National Academy of Sciences}, 99\penalty0
  (7):\penalty0 4147--4151, April 2002.
\newblock ISSN 0027-8424, 1091-6490.
\newblock \doi{10.1073/pnas.062036699}.
\newblock URL \url{https://www.pnas.org/content/99/7/4147}.

\bibitem[Kokot et~al.(2017)Kokot, Kolmakov, Aranson, and
  Snezhko]{kokot_dynamic_2017}
Gašper Kokot, German~V. Kolmakov, Igor~S. Aranson, and Alexey Snezhko.
\newblock Dynamic self-assembly and self-organized transport of magnetic
  micro-swimmers.
\newblock \emph{Scientific Reports}, 7\penalty0 (1):\penalty0 14726, November
  2017.
\newblock ISSN 2045-2322.
\newblock \doi{10.1038/s41598-017-15193-z}.
\newblock URL \url{https://www.nature.com/articles/s41598-017-15193-z}.

\bibitem[Tierno et~al.(2008)Tierno, Golestanian, Pagonabarraga, and
  Sagués]{tierno_controlled_2008}
Pietro Tierno, Ramin Golestanian, Ignacio Pagonabarraga, and Francesc Sagués.
\newblock Controlled {Swimming} in {Confined} {Fluids} of {Magnetically}
  {Actuated} {Colloidal} {Rotors}.
\newblock \emph{Physical Review Letters}, 101\penalty0 (21):\penalty0 218304,
  November 2008.
\newblock \doi{10.1103/PhysRevLett.101.218304}.
\newblock URL \url{https://link.aps.org/doi/10.1103/PhysRevLett.101.218304}.

\bibitem[Snezhko and Aranson(2011)]{snezhko_magnetic_2011}
Alexey Snezhko and Igor~S. Aranson.
\newblock Magnetic manipulation of self-assembled colloidal asters.
\newblock \emph{Nature Materials}, 10\penalty0 (9):\penalty0 698--703,
  September 2011.
\newblock ISSN 1476-4660.
\newblock \doi{10.1038/nmat3083}.
\newblock URL \url{https://www.nature.com/articles/nmat3083}.

\bibitem[Burns et~al.(1989)Burns, Fournier, and
  Golovchenko]{burns_optical_1989}
Michael~M. Burns, Jean-Marc Fournier, and Jene~A. Golovchenko.
\newblock Optical binding.
\newblock \emph{Physical Review Letters}, 63\penalty0 (12):\penalty0
  1233--1236, September 1989.
\newblock \doi{10.1103/PhysRevLett.63.1233}.
\newblock URL \url{https://link.aps.org/doi/10.1103/PhysRevLett.63.1233}.

\bibitem[Thirunamachandran(1980)]{thirunamachandran_intermolecular_1980}
T.~Thirunamachandran.
\newblock Intermolecular interactions in the presence of an intense radiation
  field.
\newblock \emph{Molecular Physics}, 40\penalty0 (2):\penalty0 393--399, June
  1980.
\newblock ISSN 0026-8976.
\newblock \doi{10.1080/00268978000101561}.
\newblock URL \url{https://doi.org/10.1080/00268978000101561}.

\bibitem[Bai et~al.(2018)Bai, Wu, and Ge]{bai_lateral_2018}
Jing Bai, Zhen-Sen Wu, and Cheng-Xian Ge.
\newblock Lateral {Optical} {Binding} of {Stratified} {Biological} {Cells}
  {Induced} by {High}-{Order} {Bessel} {Beams}.
\newblock In \emph{2018 12th {International} {Symposium} on {Antennas},
  {Propagation} and {EM} {Theory} ({ISAPE})}, pages 1--3, December 2018.
\newblock \doi{10.1109/ISAPE.2018.8634151}.

\bibitem[Li et~al.(2021)Li, Liu, Lin, Ng, and Chan]{li_non-hermitian_2021}
Xiao Li, Yineng Liu, Zhifang Lin, Jack Ng, and C.~T. Chan.
\newblock Non-{Hermitian} physics for optical manipulation uncovers inherent
  instability of large clusters.
\newblock \emph{Nature Communications}, 12\penalty0 (1):\penalty0 6597,
  November 2021.
\newblock ISSN 2041-1723.
\newblock \doi{10.1038/s41467-021-26732-8}.
\newblock URL \url{https://www.nature.com/articles/s41467-021-26732-8}.

\bibitem[Dapasse and Vigoureux(1994)]{dapasse_optical_1994}
F~Dapasse and J~M Vigoureux.
\newblock Optical binding force between two {Rayleigh} particles.
\newblock \emph{Journal of Physics D: Applied Physics}, 27\penalty0
  (5):\penalty0 914--919, May 1994.
\newblock ISSN 0022-3727, 1361-6463.
\newblock \doi{10.1088/0022-3727/27/5/006}.
\newblock URL
  \url{https://iopscience.iop.org/article/10.1088/0022-3727/27/5/006}.

\bibitem[Mohanty et~al.(2004)Mohanty, Andrews, and Gupta]{mohanty_optical_2004}
Samarendra~Kumar Mohanty, Joseph~Thomas Andrews, and Pradeep~Kumar Gupta.
\newblock Optical binding between dielectric particles.
\newblock \emph{Optics Express}, 12\penalty0 (12):\penalty0 2746--2753, June
  2004.
\newblock ISSN 1094-4087.
\newblock \doi{10.1364/OPEX.12.002746}.
\newblock URL
  \url{https://www.osapublishing.org/oe/abstract.cfm?uri=oe-12-12-2746}.

\bibitem[Tatarkova et~al.(2002)Tatarkova, Carruthers, and
  Dholakia]{tatarkova_one-dimensional_2002}
S.~A. Tatarkova, A.~E. Carruthers, and K.~Dholakia.
\newblock One-{Dimensional} {Optically} {Bound} {Arrays} of {Microscopic}
  {Particles}.
\newblock \emph{Physical Review Letters}, 89\penalty0 (28):\penalty0 283901,
  December 2002.
\newblock \doi{10.1103/PhysRevLett.89.283901}.
\newblock URL \url{https://link.aps.org/doi/10.1103/PhysRevLett.89.283901}.

\bibitem[Ng et~al.(2005)Ng, Lin, Chan, and Sheng]{ng_photonic_2005}
Jack Ng, Z.~F. Lin, C.~T. Chan, and Ping Sheng.
\newblock Photonic clusters formed by dielectric microspheres: {Numerical}
  simulations.
\newblock \emph{Physical Review B}, 72\penalty0 (8):\penalty0 085130, August
  2005.
\newblock \doi{10.1103/PhysRevB.72.085130}.
\newblock URL \url{https://link.aps.org/doi/10.1103/PhysRevB.72.085130}.

\bibitem[Metzger et~al.(2006)Metzger, Dholakia, and
  Wright]{metzger_observation_2006}
N.~K. Metzger, K.~Dholakia, and E.~M. Wright.
\newblock Observation of {Bistability} and {Hysteresis} in {Optical} {Binding}
  of {Two} {Dielectric} {Spheres}.
\newblock \emph{Physical Review Letters}, 96\penalty0 (6):\penalty0 068102,
  February 2006.
\newblock \doi{10.1103/PhysRevLett.96.068102}.
\newblock URL \url{https://link.aps.org/doi/10.1103/PhysRevLett.96.068102}.

\bibitem[Karásek et~al.(2008)Karásek, Čižmár, Brzobohatý, Zemánek,
  Garcés-Chávez, and Dholakia]{karasek_long-range_2008}
V.~Karásek, T.~Čižmár, O.~Brzobohatý, P.~Zemánek, V.~Garcés-Chávez, and
  K.~Dholakia.
\newblock Long-{Range} {One}-{Dimensional} {Longitudinal} {Optical} {Binding}.
\newblock \emph{Physical Review Letters}, 101\penalty0 (14):\penalty0 143601,
  October 2008.
\newblock \doi{10.1103/PhysRevLett.101.143601}.
\newblock URL \url{https://link.aps.org/doi/10.1103/PhysRevLett.101.143601}.

\bibitem[Dholakia and Zemánek(2010)]{dholakia_colloquium_2010}
Kishan Dholakia and Pavel Zemánek.
\newblock Colloquium: {Gripped} by light: {Optical} binding.
\newblock \emph{Reviews of Modern Physics}, 82\penalty0 (2):\penalty0
  1767--1791, June 2010.
\newblock \doi{10.1103/RevModPhys.82.1767}.
\newblock URL \url{https://link.aps.org/doi/10.1103/RevModPhys.82.1767}.

\bibitem[Rodríguez et~al.(2008)Rodríguez, Dávila~Romero, and
  Andrews]{rodriguez_optical_2008}
Justo Rodríguez, Luciana~C. Dávila~Romero, and David~L. Andrews.
\newblock Optical binding in nanoparticle assembly: {Potential} energy
  landscapes.
\newblock \emph{Physical Review A}, 78\penalty0 (4):\penalty0 043805, October
  2008.
\newblock ISSN 1050-2947, 1094-1622.
\newblock \doi{10.1103/PhysRevA.78.043805}.
\newblock URL \url{https://link.aps.org/doi/10.1103/PhysRevA.78.043805}.

\bibitem[Forbes et~al.(2020)Forbes, Bradshaw, and Andrews]{forbes_optical_2020}
Kayn~A. Forbes, David~S. Bradshaw, and David~L. Andrews.
\newblock Optical binding of nanoparticles.
\newblock \emph{Nanophotonics}, 9\penalty0 (1):\penalty0 1--17, January 2020.
\newblock ISSN 2192-8614.
\newblock \doi{10.1515/nanoph-2019-0361}.
\newblock URL
  \url{https://www.degruyter.com/document/doi/10.1515/nanoph-2019-0361/html}.

\bibitem[Nan et~al.(2018)Nan, Han, Scherer, and Yan]{nan_dissipative_2018}
Fan Nan, Fei Han, Norbert~F. Scherer, and Zijie Yan.
\newblock Dissipative {Self}‐{Assembly} of {Anisotropic} {Nanoparticle}
  {Chains} with {Combined} {Electrodynamic} and {Electrostatic} {Interactions}.
\newblock \emph{Advanced Materials}, 30\penalty0 (45):\penalty0 1803238,
  November 2018.
\newblock ISSN 0935-9648, 1521-4095.
\newblock \doi{10.1002/adma.201803238}.
\newblock URL \url{https://onlinelibrary.wiley.com/doi/10.1002/adma.201803238}.

\bibitem[Brzobohatý et~al.(2020)Brzobohatý, Chvátal, Šiler, and
  Zemánek]{brzobohaty_complex_2020}
Oto Brzobohatý, Lukáš Chvátal, Martin Šiler, and Pavel Zemánek.
\newblock Complex colloidal structures with non-linear optical properties
  formed in an optical trap.
\newblock \emph{Optics Express}, 28\penalty0 (25):\penalty0 37700, December
  2020.
\newblock ISSN 1094-4087.
\newblock \doi{10.1364/OE.405839}.
\newblock URL \url{https://opg.optica.org/abstract.cfm?URI=oe-28-25-37700}.

\bibitem[Yan et~al.(2014)Yan, Gray, and Scherer]{yan_potential_2014}
Zijie Yan, Stephen~K. Gray, and Norbert~F. Scherer.
\newblock Potential energy surfaces and reaction pathways for light-mediated
  self-organization of metal nanoparticle clusters.
\newblock \emph{Nature Communications}, 5\penalty0 (1):\penalty0 3751,
  September 2014.
\newblock ISSN 2041-1723.
\newblock \doi{10.1038/ncomms4751}.
\newblock URL \url{http://www.nature.com/articles/ncomms4751}.

\bibitem[Draine and Flatau(1994)]{draine_discrete-dipole_1994}
Bruce~T. Draine and Piotr~J. Flatau.
\newblock Discrete-{Dipole} {Approximation} {For} {Scattering} {Calculations}.
\newblock \emph{JOSA A}, 11\penalty0 (4):\penalty0 1491--1499, April 1994.
\newblock ISSN 1520-8532.
\newblock \doi{10.1364/JOSAA.11.001491}.
\newblock URL
  \url{https://www.osapublishing.org/josaa/abstract.cfm?uri=josaa-11-4-1491}.

\bibitem[Hoekstra et~al.(2001)Hoekstra, Frijlink, Waters, and
  Sloot]{hoekstra_radiation_2001}
A.~G. Hoekstra, M.~Frijlink, L.~B. F.~M. Waters, and P.~M.~A. Sloot.
\newblock Radiation forces in the discrete-dipole approximation.
\newblock \emph{JOSA A}, 18\penalty0 (8):\penalty0 1944--1953, August 2001.
\newblock ISSN 1520-8532.
\newblock \doi{10.1364/JOSAA.18.001944}.
\newblock URL
  \url{https://www.osapublishing.org/josaa/abstract.cfm?uri=josaa-18-8-1944}.

\bibitem[Ashkin et~al.(1986)Ashkin, Dziedzic, Bjorkholm, and
  Chu]{ashkin_observation_1986}
A.~Ashkin, J.~M. Dziedzic, J.~E. Bjorkholm, and Steven Chu.
\newblock Observation of a single-beam gradient force optical trap for
  dielectric particles.
\newblock \emph{Optics Letters}, 11\penalty0 (5):\penalty0 288--290, May 1986.
\newblock ISSN 1539-4794.
\newblock \doi{10.1364/OL.11.000288}.
\newblock URL
  \url{https://www.osapublishing.org/ol/abstract.cfm?uri=ol-11-5-288}.

\bibitem[Wei et~al.(2016)Wei, Ng, Chan, and Ou-Yang]{wei_lateral_2016}
Ming-Tzo Wei, Jack Ng, C.~T. Chan, and H.~Daniel Ou-Yang.
\newblock Lateral optical binding between two colloidal particles.
\newblock \emph{Scientific Reports}, 6\penalty0 (1):\penalty0 38883, December
  2016.
\newblock ISSN 2045-2322.
\newblock \doi{10.1038/srep38883}.
\newblock URL \url{https://www.nature.com/articles/srep38883}.

\bibitem[Karásek et~al.(2006)Karásek, Dholakia, and
  Zemánek]{karasek_analysis_2006}
V.~Karásek, K.~Dholakia, and P.~Zemánek.
\newblock Analysis of optical binding in one dimension.
\newblock \emph{Applied Physics B}, 84\penalty0 (1):\penalty0 149--156, July
  2006.
\newblock ISSN 1432-0649.
\newblock \doi{10.1007/s00340-006-2297-8}.
\newblock URL \url{https://doi.org/10.1007/s00340-006-2297-8}.

\bibitem[Karásek et~al.(2009)Karásek, Brzobohatý, and
  Zemánek]{karasek_longitudinal_2009}
V~Karásek, O~Brzobohatý, and P~Zemánek.
\newblock Longitudinal optical binding of several spherical particles studied
  by the coupled dipole method.
\newblock \emph{Journal of Optics A: Pure and Applied Optics}, 11\penalty0
  (3):\penalty0 034009, March 2009.
\newblock ISSN 1464-4258, 1741-3567.
\newblock \doi{10.1088/1464-4258/11/3/034009}.
\newblock URL
  \url{https://iopscience.iop.org/article/10.1088/1464-4258/11/3/034009}.

\bibitem[Singer et~al.(2003)Singer, Frick, Bernet, and
  Ritsch-Marte]{singer_self-organized_2003}
Wolfgang Singer, Manfred Frick, Stefan Bernet, and Monika Ritsch-Marte.
\newblock Self-organized array of regularly spaced microbeads in a
  fiber-optical trap.
\newblock \emph{JOSA B}, 20\penalty0 (7):\penalty0 1568--1574, July 2003.
\newblock ISSN 1520-8540.
\newblock \doi{10.1364/JOSAB.20.001568}.
\newblock URL
  \url{https://www.osapublishing.org/josab/abstract.cfm?uri=josab-20-7-1568}.

\bibitem[Zhang et~al.(2020)Zhang, Qiu, Qiu, Qiu, Li, Li, Zhao, Zhao, Liu, and
  Liu]{zhang_complex_2020}
Xiaoning Zhang, Jun Qiu, Jun Qiu, Jun Qiu, Xingcan Li, Xingcan Li, Junming
  Zhao, Junming Zhao, Linhua Liu, and Linhua Liu.
\newblock Complex refractive indices measurements of polymers in visible and
  near-infrared bands.
\newblock \emph{Applied Optics}, 59\penalty0 (8):\penalty0 2337--2344, March
  2020.
\newblock ISSN 2155-3165.
\newblock \doi{10.1364/AO.383831}.
\newblock URL
  \url{https://www.osapublishing.org/ao/abstract.cfm?uri=ao-59-8-2337}.

\bibitem[Hale and Querry(1973)]{hale_optical_1973}
G.~M. Hale and M.~R. Querry.
\newblock Optical {Constants} of {Water} in the 200-nm to 200-microm
  {Wavelength} {Region}.
\newblock \emph{Applied Optics}, 12\penalty0 (3):\penalty0 555--563, March
  1973.
\newblock ISSN 1559-128X.
\newblock \doi{10.1364/AO.12.000555}.

\bibitem[Malitson(1965)]{malitson_interspecimen_1965}
I.~H. Malitson.
\newblock Interspecimen {Comparison} of the {Refractive} {Index} of {Fused}
  {Silica}*,†.
\newblock \emph{JOSA}, 55\penalty0 (10):\penalty0 1205--1209, October 1965.
\newblock \doi{10.1364/JOSA.55.001205}.
\newblock URL
  \url{https://www.osapublishing.org/josa/abstract.cfm?uri=josa-55-10-1205}.

\bibitem[Crocker and Grier(1996)]{crocker_methods_1996}
John~C. Crocker and David~G. Grier.
\newblock Methods of {Digital} {Video} {Microscopy} for {Colloidal} {Studies}.
\newblock \emph{Journal of Colloid and Interface Science}, 179\penalty0
  (1):\penalty0 298--310, April 1996.
\newblock ISSN 0021-9797.
\newblock \doi{10.1006/jcis.1996.0217}.
\newblock URL
  \url{https://www.sciencedirect.com/science/article/pii/S0021979796902179}.

\bibitem[Allan et~al.(2015)Allan, Caswell, Keim, and van~der
  Wel]{allan_trackpy_2015}
Dan Allan, Thomas Caswell, Nathan Keim, and Casper van~der Wel.
\newblock trackpy: {Trackpy} v0.3.0, November 2015.
\newblock URL \url{https://zenodo.org/record/34028}.

\end{thebibliography}


\begin{thebibliography}{5}
\providecommand{\natexlab}[1]{#1}
\providecommand{\url}[1]{\texttt{#1}}
\expandafter\ifx\csname urlstyle\endcsname\relax
  \providecommand{\doi}[1]{doi: #1}\else
  \providecommand{\doi}{doi: \begingroup \urlstyle{rm}\Url}\fi

\bibitem[Ashkin()]{ashkin_acceleration_1970}
A.~Ashkin.
\newblock Acceleration and trapping of particles by radiation pressure.
\newblock \emph{Physical Review Letters}, 24\penalty0 (4):\penalty0 156--159.
\newblock \doi{10.1103/PhysRevLett.24.156}.
\newblock URL \url{https://link.aps.org/doi/10.1103/PhysRevLett.24.156}.

\bibitem[Ashkin et~al.()Ashkin, Dziedzic, Bjorkholm, and
  Chu]{ashkin_observation_1986}
A.~Ashkin, J.~M. Dziedzic, J.~E. Bjorkholm, and Steven Chu.
\newblock Observation of a single-beam gradient force optical trap for
  dielectric particles.
\newblock \emph{Optics Letters}, 11\penalty0 (5):\penalty0 288--290.
\newblock ISSN 1539-4794.
\newblock \doi{10.1364/OL.11.000288}.
\newblock URL
  \url{https://www.osapublishing.org/ol/abstract.cfm?uri=ol-11-5-288}.

\bibitem[Burns et~al.()Burns, Fournier, and Golovchenko]{burns_optical_1989}
Michael~M. Burns, Jean-Marc Fournier, and Jene~A. Golovchenko.
\newblock Optical binding.
\newblock \emph{Physical Review Letters}, 63\penalty0 (12):\penalty0
  1233--1236.
\newblock \doi{10.1103/PhysRevLett.63.1233}.
\newblock URL \url{https://link.aps.org/doi/10.1103/PhysRevLett.63.1233}.

\bibitem[Hoekstra et~al.()Hoekstra, Frijlink, Waters, and
  Sloot]{hoekstra_radiation_2001}
A.~G. Hoekstra, M.~Frijlink, L.~B. F.~M. Waters, and P.~M.~A. Sloot.
\newblock Radiation forces in the discrete-dipole approximation.
\newblock \emph{{JOSA} A}, 18\penalty0 (8):\penalty0 1944--1953.
\newblock ISSN 1520-8532.
\newblock \doi{10.1364/JOSAA.18.001944}.
\newblock URL
  \url{https://www.osapublishing.org/josaa/abstract.cfm?uri=josaa-18-8-1944}.

\bibitem[Draine and Flatau()]{draine_discrete-dipole_1994}
Bruce~T. Draine and Piotr~J. Flatau.
\newblock Discrete-dipole approximation for scattering calculations.
\newblock \emph{{JOSA} A}, 11\penalty0 (4):\penalty0 1491--1499.
\newblock ISSN 1520-8532.
\newblock \doi{10.1364/JOSAA.11.001491}.
\newblock URL
  \url{https://www.osapublishing.org/josaa/abstract.cfm?uri=josaa-11-4-1491}.

\end{thebibliography}

\end{document}


\maketitle


\large
\section{Optical scattering, gradient, and binding forces}

The analysis in the manuscript was guided by comparing two types of scattering bodies, Polystyrene (PS) and Silica (SiO$_2$) particles, denoted as HIP and LIP.
Initial calculations were done comparing the relative strengths of three optical forces: the scattering force, gradient force, and binding force.

The scattering force is a first-order optical force which arises from the back-reflection of incident photons \cite{ashkin_acceleration_1970}.
The scattering force is non-conservative and is dependent on the direction of light propagation.
We determine the strength of the force through simulating a single particle in single plane wave.

The gradient force is a first-order optical force due to the induced dipole moments generated in the scattering object by the incident light \cite{ashkin_observation_1986}.
The force is completely conservative and is highly depended on the gradients in the electric field.
We determine the strength of the force through simulating a single particle in a standing wave generated by two counter-propagating plane waves.
The two counter-propagating waves negate scattering forces, the standing wave pattern generates gradients in the optical field which apply gradient forces onto the single object.

The optical binding force is a second-order optical force due to the modification of an incident field on an object by the scattered fields of other objects \cite{burns_optical_1989}. 
Optical binding is a combination of conservative and non-conservative forces.
We simulate the force by placing two objects separated in cross-polarized counter-propagating plane waves.
The counter-propagation negates the first-order scattering forces while cross-polarizing removes the standing wave suppressing the first-order gradient force.
Remaining forces must be due to optical binding effects.

At the size range of interest we find that PS particles have relatively stronger scattering forces while LIPs have relatively stronger gradient forces.
The conclusion made is that HIPs are stronger scatterers.

\section{Numerical Simulations}

The numerical simulations were achieved using a Coupled Dipole Method (CDM) based scattering code \cite{hoekstra_radiation_2001,draine_discrete-dipole_1994}.
In each case, we ensure 10 dipoles per wavelength is used.
We use an iterative solver in which the number of iterations are set until the root mean square of the dipole moments converge to $< 1 \%$.

The 2-dimensional maps were generated by running a stand-alone simulation at each particle displacement.
A single particle was fixed at an origin while a second particle was moved along a 2-dimensional space on the plane in the immediate vicinity of the fixed particle.

\section{Force map generation}

The force maps were generated using CDM based simulation.
By placing two particles in a cross-polarized counter-propagating field, we calculate the radiation forces placed on each particle.
Filling out the 2-dimensional space is done by fixing one object at an origin and repeating the simulation for various displacements.

\section{Correlating volume mean squared difference (VMSD) with mean squared displacement (MSD)}

\begin{figure*}[t!]
    \centering
    \includegraphics{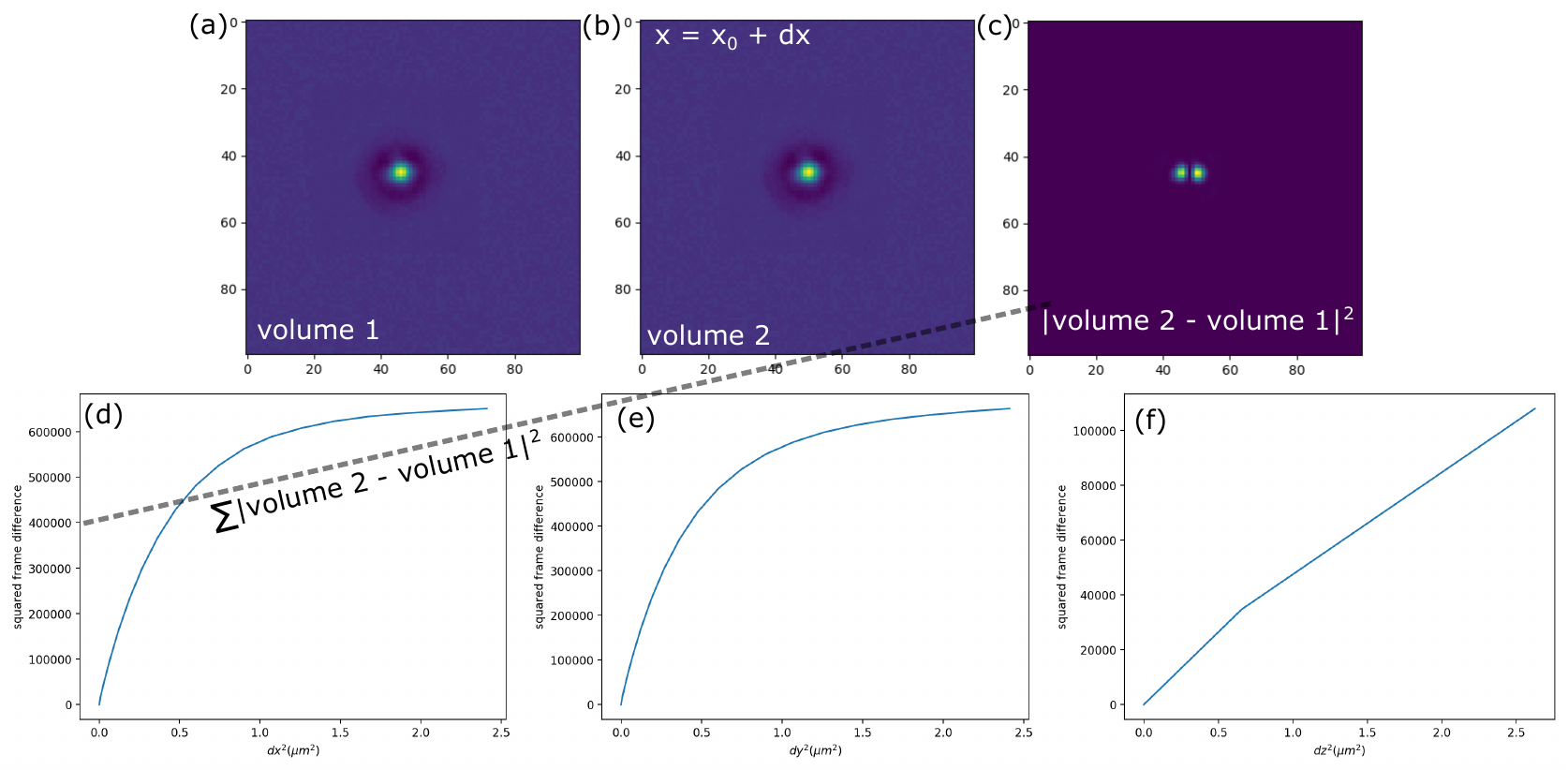}
    \caption{A simulated PS particle displaced along the x, y, and z axis plotted against a corresponding squared volume difference.}
    \label{fig:ref_fmsd}
\end{figure*}

\begin{figure}[h]
    \centering
    \includegraphics{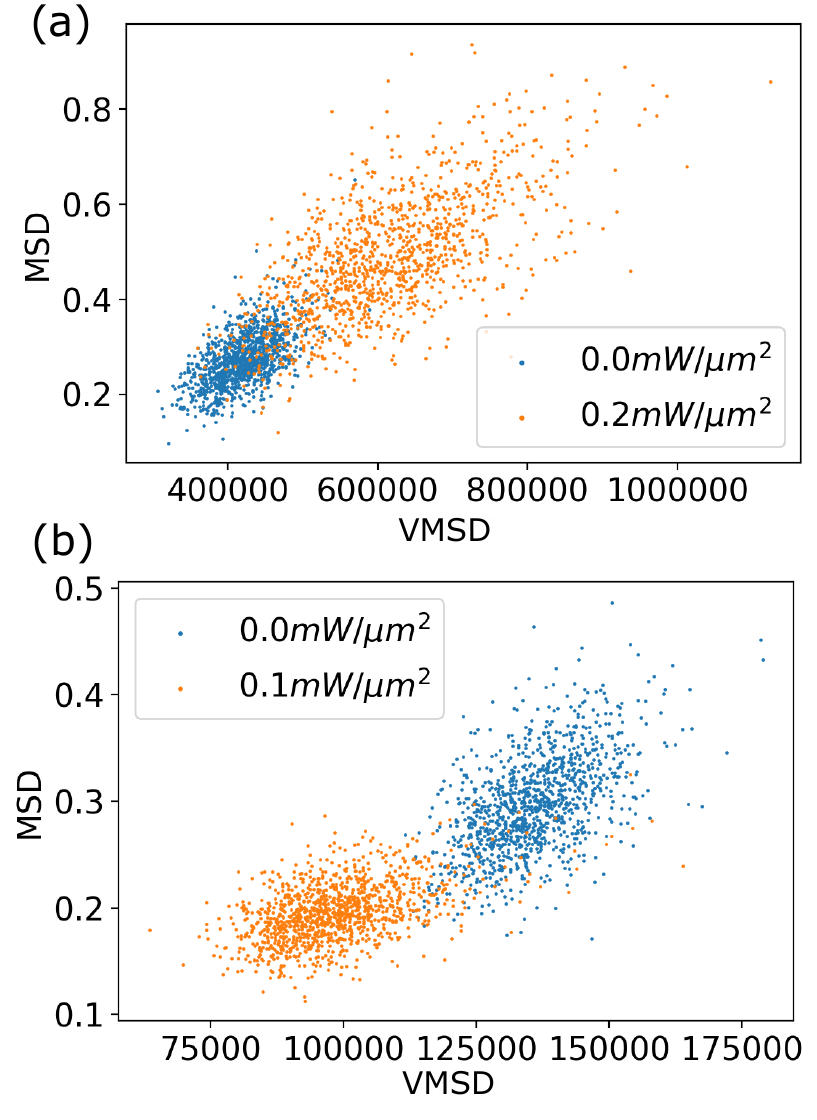}
    \caption{VMSD values compared to 2-dimensional mean squared displacement (MSD) in the xy-plane for (a) HIP (PS) and (b) LIP (SiO$_2$) particles.}
    \label{fig:msd_VMSD}
\end{figure}

3-Dimensional videos were taken of particles undergoing Brownian motion.
19 individual particles located in a single volume were averaged to generate a reference particle.
We used the reference particle to simulate particle motion by shifting it a known amount in a generated noisy background.
The results of the simulated displacements along different axes are provided in \figref{ref_fmsd}.
We can correlate the size of the simulated step to the value of taking the sum squared volume difference (VSD) of the array before and after the step.
We found that larger displacements correspond to a larger VSD in general.
One limitation is that the motion in the z-direction is underrepresented compared to motion in the x-y plane due to the inherent asymmetry of the particles imaging profile along the z-axis.
This limitation does not allow us to place an absolute estimate of how fast the objects are moving in physical units; however we can still measure the relative motion under the assumption that the motion in the compared systems share similar axial preferences.
Observations of the data support this assumption.
The second limitation is that the VSD value is capped at displacements that are slightly larger than the particle diameter.
At the volume rate of 10 volumes/s, particles undergoing Brownian motion are moving slow enough; however we have observed that particles undergoing optical binding forces can be propelled much faster.
One way to alleviate the limitation is to record at a higher volume rate; however, it is currently unfeasible to cover the full range of possible particle speeds.
This limitation is also present in traditional tracking methods.

Although the motion of the fastest particles are underrepresented in the VSD values, the VMSD method still allows a maximum value to be assigned.
Traditional tracking methods would not be able to assign a velocity value at all.
We also note that underrepresenting the motion of the faster moving particles only acts to underrepresent the strength of the findings.

The VMSD at any given time is determined by subtracting two subsequent volumes from each other, removing the contribution from the background, and dividing by the total number of particles in the frame:
\begin{equation}
    \Delta(t) = \frac{1}{N_p(t)} \left[\sum_{i,j,k}^{volume}(p_{i,j,k,t} - p_{i,j,k,t-dt})^2  - \Delta_{0}\right]
\end{equation}
The background contribution, $\Delta_{0}$, is determined in each video by monitoring the squared difference of sections that have zero particles.
We find a non-zero constant value which can be subtracted to remove contributions that are due to noise.
The number of particles, $N_p$, is estimated by the following:
\begin{equation}
    N_p = c_0 \sum_{i,j,k}^{volume}{|p_{i,j,k,t} - b_{i,j,k}|} + c_1
\end{equation}
Where $b_{i,j,k}$ is belongs to a volume representing the sample devoid of particles.
$c_0$ and $c_1$ are fitting parameters determined by linearly fitting the number of particles found by traditional tracking methods to $\sum_{i,j,k}^{volume}{|p_{i,j,k,t} - b_{i,j,k}|}$ for a system of particles undergoing Brownian motion.
Some results in \figref{msd_VMSD}.
For the HIPs, we show that for diffusive particles and particles interacting in a weak field, there is a clear correlation between the VMSD and the true MSD (\figref{msd_VMSD}).
For the LIPs, we also find correlation between the VMSD and true MSD for the diffusive particles; however in a weak field, the correlation is reduced.
We found that in a weak field, the Lower-index particles tended to pack into tight formations which effected the individual tracking capabilities.
Thus we utilize the purely diffusive correlation to justify the use of VMSD in a difficult tracking atmosphere.

We represent the motion in terms of the VMSD values relative to the Brownian Motion case.
Doing so allows us to compare the HIP and LIP systems in how their motion changes relative to diffusive motion.

\section{Electronic Supplementary Information}

All movies are 2 mins played in real-time.

\subsection{Movie S1}

HIP (Polystyrene) particles in cross polarized counter-propagating beam (peak power density - 1.5 mW/$\mu m^2$).

\subsection{Movie S2}

LIP (SiO$_2$) particles in cross polarized counter-propagating beam (peak power density - 1.5 mW/$\mu m^2$).

\subsection{Movie S2}

HIP (Polystyrene) particles in standing wave (peak power density - 1.5 mW/$\mu m^2$).

\bibliographystyle{unsrtnat}
\bibliography{supplement}